\documentclass[sigconf,nonacm,10pt]{acmart}
\pdfoutput=1

\usepackage{booktabs} % for professional tables
\usepackage[capitalize,noabbrev]{cleveref}
\usepackage{makecell}
\usepackage{tikzsymbols}

\begin{document}

\title[Human-Human Pair Programming vs. Human-AI pAIr Programming]{{ Is AI the better programming partner? \\ Human-Human Pair Programming vs. Human-AI pAIr Programming}}

\author{Qianou Ma}
\authornote{Corresponding author}
\affiliation{%
  \institution{Carnegie Mellon University}
  \city{Pittsburgh}
  \country{USA}
}
\email{qianouma@cmu.edu}

\author{Tongshuang Wu} 
\affiliation{%
  \institution{Carnegie Mellon University}
  \city{Pittsburgh}
  \country{USA}
}
\email{sherryw@cs.cmu.edu}

\author{Kenneth Koedinger} 
\affiliation{%
  \institution{Carnegie Mellon University}
  \city{Pittsburgh}
  \country{USA}
}
\email{koedinger@cmu.edu}


\begin{abstract}

The emergence of large-language models (LLMs) that excel at code generation and commercial products such as GitHub's Copilot has sparked interest in human-AI pair programming (referred to as ``pAIr programming'') where an AI system collaborates with a human programmer. 
While traditional pair programming between humans has been extensively studied, it remains uncertain whether its findings can be applied to human-AI pair programming.
We compare human-human and human-AI pair programming, exploring their similarities and differences in interaction, measures, benefits, and challenges.
We find that the effectiveness of both approaches is mixed in the literature (though the measures used for pAIr programming are not as comprehensive). 
We summarize moderating factors on the success of human-human pair programming, which provides opportunities for pAIr programming research. For example, mismatched expertise makes pair programming less productive, therefore well-designed AI programming assistants may adapt to differences in expertise levels.
\end{abstract}


\maketitle

\section{Introduction}
\label{intro}

Pair programming was first introduced in the 1990s as part of the Agile software development practice~\cite{Beck1999-hm}. In its original definition, pair programming describes the practice of two programmers working together on the same task using a single computer, keyboard, and mouse. One programmer in the pair, the ``driver,'' performs the coding (typing) and implements the task, while the other programmer, the ``navigator,'' aids in planning, reviewing, debugging, and suggesting improvements and alternatives. % \cite{Beck1999-hm}. 
Over time, pair programming has evolved and adapted to different contexts and purposes. Now, it is used in a wide range of settings, including education, industry, and open-source software development~\cite{Alves_De_Lima_Salge2016-se, Umapathy2017-cd}. 

Recent advances in code-generating large-language models (LLMs) have led to the widespread popularity of commercial AI-powered programming assistance tools such as GitHub Copilot \cite{GitHub2021-kz}, which advertises itself as ``your AI pair programmer.'' For pAIr programming, instead of two humans working on a single computer, it is the programmer and the LLM-based AI that work together on the same task.
The shift in the paradigm raises the questions: \emph{Is the AI programming partner comparable to a human pair programmer? Are they applicable to the same contexts, can they achieve similar or better performance, and should people interact with them in the same way?} 

In this work, we delve into the current state of research on human-human and human-AI pair programming to uncover their similarities and differences, and we hope to inspire better evaluations and designs of code-generating LLMs as a pAIr programmer.
We start by reviewing the application context, methods, and tasks for both human-human and human-AI pair programming literature (\cref{context}), then dive into fine-grained comparisons of their measurements of success (\cref{mixed-outcome}), as well as the contributing moderators, e.g., pair compatibility factors like expertise (\cref{moderator}). 

We find that (1) prior work on both pair programming paradigms has observed mixed results in \emph{quality}, \emph{productivity}, \emph{satisfaction}, \emph{learning}, and \emph{cost}, (2) pAIr programming has yet to develop comprehensive measurements, and (3) key factors to pAIr's success have been largely unexplored. 

Building on our exploration, we then discuss views and challenges of characterizing AI as a pair programmer, and elaborate on future opportunities for developing best practices and guidelines for human-AI pair programming (\cref{discussion}). 
First, we argue that moderating factors that bring challenges to human-human pair programming (e.g., compatibility and communication) unveil opportunities to improve human-AI pair programming. It can be promising to exploit the differences between a human and an AI partner (e.g., more customizable expertise level and more adaptable communication styles) to design for more successful human-AI pair programming experiences.
Second, we encourage future research to explore the best deployment environment for human-AI pair programming. 
While most human-AI pair programming works have focused on assisting professional developers, 
we hope 
to inspire more future works in the learning context (or, student-AI pair programming), and we highlight potential challenges involved.

\section{Contexts, Methods, and Tasks}
\label{context}

Human-human pair programming originated as a practice in the software engineering industry \cite{Beck1999-hm} and then become a popular collaborative learning practice in classrooms \cite{Umapathy2017-cd}. 
Therefore, in this paper, we compare human-human and human-AI pair programming in both the industry and education contexts, as they are the most common contexts.

We adhere to the original definition of human-human pair programming to closely resemble human-AI interaction on a single device. Other modes of interaction exist for comparing human and human-AI teams in programming tasks, such as computer-mediated collaborative learning \cite{Sankaranarayanan2020-fc} and distributed pair programming \cite{Da_Silva_Estacio2015-nh}, but they are beyond the scope of this paper. 

For human-AI pair programming, most current works have been evaluating Copilot using case studies (e.g., \cite{Bird2023-jv}) or experimental studies (e.g., \cite{Vaithilingam2022-jh}) with experienced programmers in the industry. Similar to human-human pair programming, researchers tried to mimic a realistic professional development environment in their task setup. For example, \citet{Barke2022-tv} invited 20 participants, mostly doctoral students and software engineers, to complete tasks such as developing Chat Client and Server. However, there is a lack of non-invasive field observation studies like what human-human pair programming studies have done \cite{Sillitti2012-ma, Plonka2011-vd}.

Few recent works have explored using LLM-based programming environments or Copilot with students. For example, \citet{Kazemitabaar2023-gu} used a controlled experimental study with 69 novice students from 10 to 17 years old working on 45 Python code-authoring and code-modifying tasks. 
However, existing works on human-AI pair programming are mostly in lab experiments, and there is still a lack of large-scale study \cite{McDowell2002-bi} and classroom deployment \cite{Nagappan2003-hb, Williams2002-de} as in the human-human pair programming literature.

When setting up comparison groups, existing pAIr programming works have been comparing the human-AI pair against human-human \cite{Imai2022-so} or human solo (e.g., compare developers' work when they use Copilot or the default code completion tool) \cite{Vaithilingam2022-jh}. No current study sets up a three-way comparison for human-AI, human-human, and human-solo.

\paragraph{Summary:} In comparison to human-human pair programming works, existing pAIr studies lack realistic deployment in the workspace or classroom, and a larger sample size would also be desirable. Researchers of both pair programming
paradigms use various study designs to examine what affects the effectiveness of pair programming. In \cref{mixed-outcome} and \cref{moderator}, we compare the variables and measurements they used to further uncover what is lacking in pAIr studies.

\begin{table*}[!h]
\caption{Comparison of Outcome Variables and Moderators for Human-Human Pair Programming vs. Human-AI pAIr Programming}
\label{tab:vars}
%\vskip 0.15in
\vspace{-10pt}
\setlength{\tabcolsep}{5pt}
\fontsize{8}{8.5}\selectfont

\begin{center}

\begin{tabular}{p{0.11\textwidth} |p{0.37\textwidth} |p{0.48\textwidth}}
\toprule
\textbf{Outcomes} & \textbf{Human-Human vs. Human Solo} & \textbf{Human-AI (Copilot)} \\
\midrule
Quality 
& \makecell[{t{p{0.37\textwidth}}}]{\cChangey{2} significantly lower defect density for complex code \\ \cChangey{0} no  difference for simpler code \cite{Sison2009-mp} \\
\cChangey{2} significantly higher percentage of test cases passed \cite{Williams2001-mk}}
& 
\makecell[{t{p{0.48\textwidth}}}]{
\cChangey{-2} vs. Human-Human: more lines of code deleted in next session (lower quality) \cite{Imai2022-so} 
\\ 
\cChangey{2} vs. Human Solo: significantly improve correctness score and reduce encountered errors for novice students \cite{Kazemitabaar2023-gu}
\\
\cChangey{0} vs. Human Solo: no significant difference in task success \cite{Peng2023-uf} or task success rate in given time \cite{Vaithilingam2022-jh} }
\\
\midrule
%\abovespace
Productivity 
& \makecell[{t{p{0.37\textwidth}}}]{\cChangey{-2} significantly fewer lines of code per person hour writing simpler code, \\
\cChangey{0} no significant difference writing more complex code \cite{Sison2009-mp} \\
\cChangey{2} 29\% shorter time to complete task (pair speed advantage = 1.4) \cite{Padmanabhuni2012-pp} 
\\
% \cChangey{-2} 15\% more time together on the program \cite{Cockburn2001-ci}
}
% (1.3 to 1.8), as $100/(100-\text{percent shorter time)}$ 
% e.g., ``Relative Effort Afforded by Pairs'' \cite{Lui2006-ur} 
% 15\% decrease in productivity and no significant quality (Capers Jones)
& \makecell[{t{p{0.48\textwidth}}}]{\cChangey{2} vs. Human-Human: more lines of added code \cite{Imai2022-so} \\ 
\cChangey{2} vs. Human Solo: 55.8\% reduction in completion time \cite{Peng2023-uf}
\\
\cChangey{2} vs. Human Solo: significantly increase task completion and reduce task completion time for novice students \cite{Kazemitabaar2023-gu}
\\
\cChangey{0} vs. Human Solo: no significant difference in the task completion rate in given time \cite{Vaithilingam2022-jh} }
\\
\midrule%\abovespace
Satisfaction 
& \makecell[{t{p{0.37\textwidth}}}]{\cChangey{2} higher self-ratings of satisfaction \cite{Salleh2011-dd} \\
 \cChangey{-1} students with greater self-confidence and self-efficacy less enjoy the pair programming experience \cite{Thomas2003-ud}}
& \cChangey{2} vs. Human Solo: higher self-ratings of satisfaction \cite{Vaithilingam2022-jh, Bird2023-jv, Kalliamvakou2022-ku} 
\\
\midrule%\abovespace
Learning 
& \makecell[{t{p{0.37\textwidth}}}]{ 
\cChangey{2} higher grades, exam scores \cite{Nagappan2003-hb}, and retention \cite{McDowell2006-xz} \\ 
\cChangey{2} significantly higher gains in exam performance in female students than male students \cite{Maguire2014-ee}}
%increasing perceived gains in learning for weaker students, in pair programming
% \cChangey{1} more knowledge gains for lower-level student from higher-level partners \cite{Padmanabhuni2012-pp}}
& \cChangey{0} vs. Human Solo: no significant difference in immediate and retention post-test performance of novices, students with more prior experiences have more learning gains from AI code generator \cite{Kazemitabaar2023-gu} \\
\midrule%\abovespace
Cost 
& \makecell[{t{p{0.37\textwidth}}}]{
% \cChangey{-2} more training costs, hardware, and software costs \\ %\citet{Sun2009-ja, Cockburn2001-ci, Padberg2004-kn}
\cChangey{-2} increased management workload to match, schedule a pair, resolve collaboration conflict, assess individual contributions, etc. \cite{Ally2005-vx} \\
\cChangey{2} reduced teaching staff workload (grading one assignment from a pair) \cite{Williams2001-mk}
}
& No experiment yet. \citet{Vaithilingam2022-jh, Bird2023-jv} hypothesized that human-AI may lead to more unnecessary debugging vs. Human Solo
\\ 
\midrule
\end{tabular}

\begin{tabular}{p{0.11\textwidth}| p{0.7\textwidth} |p{0.15\textwidth}}
\textbf{Moderators} & \textbf{Human-Human vs. Human Solo} & \textbf{Human-AI (Copilot)} \\
\midrule
\makecell[tl]{Task Types \\ \& Complexity} % Task complexity of the system or task
& Complex task improve quality, simple one does not \cite{Sison2009-mp}; debugging is perceived as less enjoyable or effective than comprehension or refactoring \cite{Chaparro2005-uk}
& - \\ 
\midrule
%\abovespace
\makecell[tl]{Compatibility \\ (E.g., Expertise)}
& Random pairing led to incompatible partners and conflicts during work \cite{Nagappan2003-hb}. Expertise: improve quality more effectively if pair is similarly skilled \cite{Salleh2011-dd}; less-skilled students learn more and enjoy more \cite{Maguire2014-ee, Chaparro2005-uk}; if knowledge gap is large, less-skilled programmers may tend to be more passive and disengaged \cite{Chong2007-uz}
% ; personality conflicts is one of the top reported problems \cite{Begel2008-nm} 
& - \\
\midrule
Communication
& Conversations with intermediate-level details contribute to pair programming success \cite{Freudenberg2007-jb}; different types of discourse lead to more attempts or more debug success \cite{Murphy2010-mj}
& - \\
\midrule
Collaboration
& 
% Cooperative behavior and positive interdependence are key to pair programming success \cite{Preston2006-wi}; 
Over-reliance leads to conflicts and impedes satisfaction and learning, as work is entirely burdened on one partner \cite{Williams2002-de, Nagappan2003-hb};
% feelgood factor (satisfaction with collaboration) correlates with pair performance \cite{Muller2004-cy};
educators recommend regular role-switching to ensure equitable learning in collaboration \cite{Umapathy2017-cd}
& - \\ 
\midrule
Logistics 
& Scheduling difficulties \cite{Begel2008-nm}, teaching \& evaluating individual responsibility and accountability are important to collaboration success \cite{Preston2006-wi}, but can lead to increased management costs \cite{Ally2005-vx, Sun2009-ja}
& - \\
\bottomrule
\end{tabular}

\end{center}
\vskip -0.1in
\end{table*}

\section{Mixed Outcomes}
\label{mixed-outcome}

Literature reviews of human-human pair programming have suggested various benefits as well as mixed effects. In the industry context, according to \citet{Alves_De_Lima_Salge2016-se}, pair programming improves code quality, increases productivity, and enhances learning outcomes. However, according to \citet{Hannay2009-vv}, pair programming improves quality and shortens duration, but it increases effort, higher quality comes at the expense of considerably greater effort, and reduced completion time comes with lower quality. In the education context, pair programming brings benefits including higher quality software, student confidence in solutions, increased assignment grades, exam scores, success/passing rates in introductory courses, and retention \cite{Umapathy2017-cd, Hanks2011-ae, McDowell2006-xz}. All the reviews on human-human pair programming acknowledged that even though meta-analysis can show an overall trend and significant effect size, individual studies could report contradictory outcomes (see examples in \cref{tab:vars}). 

For human-AI pair programming, existing works mainly focus on quality, productivity, and satisfaction, and already demonstrated mixed results in quality and productivity \cite{Imai2022-so, Barke2022-tv, Vaithilingam2022-jh} (see examples in \cref{tab:vars}). 
Additionally, there is not enough research for a comprehensive review, so we cannot reach any conclusion on the effectiveness of human-AI pair programming yet. It is also hard to compare the human-human and human-AI pair programming literature, as they differ in what outcomes and measurements they adopt.

Therefore, in the top rows of \cref{tab:vars}, we listed the most common outcome variables in both literature (\emph{quality}, \emph{productivity}, \emph{satisfaction}, \emph{learning}, and \emph{cost}) and some sample work to demonstrate the mixed outcomes and various measures. We elaborate on the variety of ways to measure some of the listed outcomes as follows.

\subsection{Quality}
In human-human pair programming literature, quality can be measured using defect density, perceptual effort measure, readability, functionability, the number of test cases passed, code complexity, scores, expert opinions, etc. \cite{Sun2009-ja, Salleh2011-dd, Alves_De_Lima_Salge2016-se}.

\subsection{Productivity}
In human-human pair programming literature, duration, effort, and productivity are all types of ``efficiency'' outcomes that involve time and accomplishment. Productivity can be measured in terms of the number of completed tasks in a fixed unit of time, duration can be measured as the amount of elapsed or total time used to complete a fixed number of tasks to a certain standard, and effort can be measured as twice the duration, the person-hours required, etc. \cite{Alves_De_Lima_Salge2016-se}. We use productivity as an aggregated outcome variable of different measures, for consistency with the human-AI literature.

In current human-AI works, some measures are arguably too simplified as evaluation metrics, for example, \citet{Imai2022-so} used the number of lines of added code as the measure of productivity; however, the nature of interaction with Copilot (tab to accept suggestions) is likely to contribute to more added lines in the human-Copilot condition, and how valid would it represent the notion of productivity is questionable.

Note that some researchers have examined programmers' perceived productivity when working Copilot and found that it most strongly correlated with the general acceptance rate of AI-generated code
\cite{Ziegler2022-fz}. This is not included in \cref{tab:vars} to stay consistent with the human-human pair programming literature, as perceived productivity is a different measure than actual productivity. 

\subsection{Learning}

In human-human pair programming literature, learning can be assessed by quantitative measures such as assignment grades, exam scores, passing rate, and retention rate, or qualitative measures of higher-order thinking skills \cite{McDowell2006-xz, Umapathy2017-cd, Hanks2011-ae}.

\subsection{Cost}

In terms of cost, there is the observation that participants faced challenges in understanding and debugging Copilot's generated code, which leads to the hypothesis that human-AI pair programming could cost additional efforts and hinder programmers' task-solving effectiveness \cite{Vaithilingam2022-jh, Bird2023-jv}. 
However, \citet{Dakhel2022-he} shows that although Copilot's code could be less correct than human code, its bugs are easier to debug than human errors. 
There is currently no work that experimentally characterizes the costs of human-AI pair programming.

\paragraph{Summary:} The literature on human-human pair programming has shown mixed results in many outcome variables, including \emph{quality}, \emph{productivity}, \emph{satisfaction}, \emph{learning}, and \emph{cost}. For human-AI pair programming, or mostly human-Copilot in this paper, there are still only few works with incomprehensive measures, but a mixed outcome is also observed. We further review the potential causes of mixed outcomes of both modes of pair programming in \cref{moderator}.

\section{Moderators}
\label{moderator}

In search of the explanations of the cost-benefit of human-human pair programming experiences, researchers have found moderators such as 
\emph{task type \& complexity} \cite{Hannay2009-vv}, 
\emph{compatibility} factors like expertise \cite{Preston2006-wi, Arisholm2007-zk}, 
\emph{communication} \cite{Plonka2011-vd, Freudenberg2007-jb, Chong2007-uz}, 
\emph{collaboration} factors like over-reliance and role-switching \cite{Williams2002-de, Hannay2010-dc, Salleh2011-dd}, and
\emph{logistics} difficulties including scheduling and training \cite{Begel2008-nm, Hannay2009-vv} (as shown in the bottom rows of \cref{tab:vars}). 

These key factors influence the success of human-human pair programming. If they work well, pair programming helps programmers catch errors more easily, solve problems more quickly, review code more thoroughly, and produce overall higher-quality code; it also promotes knowledge sharing among team members, which can lead to a more cohesive and effective team. If not, challenges such as scheduling and finding suitable pairs with compatible working styles usually result in a low cost-efficiency in pair programming, and slow down the development process if there are conflicts or disagreements between pair partners \cite{Begel2008-nm, Cockburn2001-ci}.

For human-AI pair programming's moderators, much was unexplored -- we do not know what could make human-AI pair programming more or less effective. Therefore, in this section, we discuss the key moderators that are examined in the human-human pair programming literature, and individual examples of moderating effects are provided in \cref{tab:vars}.

\subsection{Task Types \& Complexity}
For task type and task complexity, \citet{Chaparro2005-uk} found that debugging tasks lead to less satisfaction and perceived efficacy compared to comprehension and refactoring tasks. 
\citet{Hannay2009-vv} found that the duration is shorter for low complexity tasks, at the expense of lower quality results, and quality is higher when complexity is higher, but it requires considerably greater effort. \citet{Arisholm2007-zk} found that the moderating effect of complexity also depends on the expertise of the pair, where ``benefits of correctness on complex system apply mainly to juniors, whereas the reductions in duration to perform the tasks correctly on the simple system apply mainly to intermediates and seniors.'' 

\subsection{Compatibility}
\citet{Salleh2011-dd} listed multiple factors for pair compatibility, such as personality, perceived skills, actual skills (expertise), self-esteem, gender, and work ethic.
% and called for more examinations on pair compatibility factors that influence effective pair formation.
\citet{Thomas2003-ud} found that paired students with similar self-confidence levels produce their best work. \citet{Hannay2010-dc} found that Big Five personality traits only have modest predictive value on pair programming performance, and expertise, task complexity, and country have stronger prediction power in comparison. There also seems to be evidence that women benefit from pair programming more than men do \cite{Preston2006-wi, Hanks2011-ae}. 

Expertise as a compatibility factor has been extensively studied in the human-human pair programming literature. For example, researchers found that a student pair performs the best when their expertise is similar \cite{Salleh2011-dd} and students preferred to be paired with similarly skilled partners \cite{Chaparro2005-uk}. However, in industry, \citet{Jensen2005-jb} reported that when both members were near the same capability level and strongly opinionated, the collaboration was counter-productive and troublesome. 

In the introductory programming context,  
\citet{Lui2006-ur} found that pairing up novices results in a larger improvement in productivity than pairing up experts. 
However, there are concerns about the risk of ``the blind leading the blind'' if they don't have an expert to consult with \cite{Ally2005-vx}. 
Researchers also found that less-skilled students learn and enjoy more than more-skilled students in pair programming \cite{Chaparro2005-uk, Maguire2014-ee}. However, when the knowledge gap is too large, students can be less satisfied and the benefits of quality may be smaller \cite{Padmanabhuni2012-pp}. \citet{Chong2007-uz} reported that a novice programmer collaborating with an expert may become disengaged, have lower self-esteem, and be afraid of slowing down or annoying their more-skilled partner \cite{Ally2005-vx}.

\subsection{Communication}
According to \citet{Freudenberg2007-jb}, ``the key to the success of pair programming [is] the proliferation of talk at an intermediate level of detail in pair programmers' conversations.'' Researchers also found that pair programming eliminates distracting activity and enables programmers to focus on productive activity \cite{Sillitti2012-ma}, which could be why engaging communications contribute to the success of pair programming. 
\citet{Murphy2010-mj} used transactive analysis to break down communication by different types of transactions, and they found that attempting more problems associated with more completion transactions and debugging success correlated with more critique transactions. Some other works pointed out the social support aspect of communication \cite{Chong2007-uz} and an explanation effect where the verbalization of the thought process makes it clearer \cite{Bird2023-jv}. 

In human-human pair programming, programmers spend about $1/3$ of the time primarily focusing on communication \cite{Plonka2011-vd}, which forces them to concentrate, rationalize, and explain their thoughts \cite{Sillitti2012-ma, Hannay2009-vv}. 
In human-AI pair programming, \citet{Mozannar2022-ng} has shown that an analogous $1/3$ amount of time is spent communicating with Copilot, such as thinking and verifying (22.4\%) Copilot's suggestion, which may be replicating the self-explanation effects in some ways, and prompt crafting, which takes 11.56\% of the time. These activities are arguably efforts to understand and communicate with Copilot. However, there is no other human to co-verify the answers, and there is no study that evaluate the communicative nature of human-Copilot interaction as human-human pair programming.

\subsection{Collaboration}

How well partners collaborate have been important factors that affect pair programming effectiveness \cite{Ally2005-vx, Sun2009-ja}, and cooperative behavior and positive interdependence are key to pair programming success \cite{Preston2006-wi}.

Collaboration can fail in various ways in a human-human pair. For example, the free-rider problem, where the entire workload is on one partner while the other remains a marginal player, can result in less satisfaction and learning \cite{Williams2002-de, Nagappan2003-hb}. In human-AI pair programming, educators are worried that easily available code-generation tools may lead to cheating, and over-reliance on AI may hinder students learning \cite{Becker2023-ue}. However, no study has formally evaluated it. 

For human-human pair programming, there is a suggested collaboration pattern of role-switching -- two software developers  periodically and regularly switch between writing code (driver) and suggesting code (navigator), aiming to ensure that both are engaged in the task and alleviate the physical and cognitive load borne by the driver \cite{Alves_De_Lima_Salge2016-se, Plonka2011-vd}.

Some researchers \citet{Freudenberg2007-jb} argue that the success of pair programming should be attributed to communication rather than ``the differences in behavior or focus between the driver and navigator,'' as they found both driver and navigator worked on similar levels of abstraction. Nevertheless, instructors still recommend drivers and navigators to regularly alternate roles to ensure equitable learning experiences \cite{Umapathy2017-cd}. 

In human-AI interaction, given Copilot's amazing capability to write code in different languages, some have argued that Copilot can take on the role of the ``driver'' in pair programming, allowing a solo programmer to take on the role of the ``navigator'' and focus on understanding the code at a higher level \cite{Imai2022-so}. However, while it is possible for humans to offload some API lookup and syntax details to Copilot, humans still need to jump back into the driver's seat frequently and fluidly switch between the thinking and writing activities \cite{Mozannar2022-ng}. It is ultimately the human programmer's sole responsibility to understand code at the statement level \cite{Sarkar2022-cy}.

\subsection{Logistics} 

Logistical challenges, including scheduling difficulties, teaching and evaluating collaboration for the pair, and figuring out individual accountability and responsibility \cite{Begel2008-nm, Preston2006-wi}, can add to the management cost of human-human pair programming \cite{Ally2005-vx, Sun2009-ja}. 

In human-AI pair programming, some may argue that the human is solely responsible in the human-AI pair \cite{Sarkar2022-cy}, but the accountability of these LLM-based generative AI is still under debate \cite{Becker2023-ue}. There may be new logistics issues for the human-AI pair, such as teaching humans how to best collaborate with Copilot. There could also be unique challenges as in every human-AI interaction scenario, such as bias, trust, and technical limitations -- much to be explored. More study would be needed to empirically and experimentally verify the moderating effects of different variables in human-AI pair programming.

\paragraph{Summary:} Human-human pair programming literature have found moderators including \emph{task type \& complexity},
\emph{compatibility}, \emph{communication}, \emph{collaboration}, and \emph{logistics}. However, there is a lack of in-depth examination of potential moderating effects in current pAIr works.

\section{Discussion and Future Work}
\label{discussion}

\subsection{LLM, Your pAIr Programmer?}

Before the occurrence of LLM-based tools that claim to be ``your AI pair programmer \cite{GitHub2021-kz},'' people have already been developing AI-powered systems to assist programmers, such as code completion tools (e.g., Tabnine), code refactoring and formal verification systems, and code synthesis and debugging tools. The evaluation focus has mostly been on usability design, cost-efficiency, and productivity \cite{Myers2016-jk, Mozannar2022-ng}, but not on the feasibility of using these AI-assisted programming tools as the pair programming partner. 

With recent advancements in generative LLM technologies, commercial AI tools like Copilot which are capable of offering real-time code suggestions and feedback beyond auto-completion seem to have a closer resemblance to a pair programming partner \cite{Bird2023-jv}. 
Many studies have evaluated and critiqued Copilot's ability to generate correct, efficient \cite{Dakhel2022-he, Nguyen2022-hk}, secure \cite{Pearce2021-et, Asare2022-pu}, readable \cite{Al_Madi2023-fg}, and verifiable \cite{Wong2022-dt} code. 
Without doubt, Copilot generates defects and errors in its suggested code, but humans are far from error-free either. A programmer cannot be and does not need to be perfect to bring benefit into the pair programming experiences, but would Copilot be qualified as a programming partner? 

In answering this question, researchers start to look into the interaction dynamics between programmers and the claimed AI pair programmer. 
Some researchers argue against the characterization of AI-assisted programming as pair programming. They believe that the analogy to human-AI pair programming is rather superficial, as what makes human-human pair programming effective (e.g., productive communication) disappear in human-AI pair programming.
According to \citet{Sarkar2022-cy}, ``LLM-assisted programming ought to be viewed as a new way of programming with its own distinct properties and challenges.'' 

We used the phrase ``human-AI pAIr programming'' in this paper, simply because we adopt the  definition of pair programming that a pair work on the same device and the same task, so we can conveniently compare human and AI as a pair programming partner.
As reviewed in \cref{mixed-outcome} and \cref{moderator}, Copilot and a human partner share a lot of similar outcomes in pair programming, but the moderators for human-AI pair programming are less examined. We believe that this comparison is meaningful in that it helps us derive insights to keep improving LLM-based programming tools.

Note that in this paper, we mostly covered studies using the VSCode Extension Copilot. Tools like ChatGPT may support the communication aspect better than Copilot \cite{Thorp2023-ah}, and there are also Bard developed by \citet{Google_undated-lx} and an experimental version of Copilot Labs by \citet{Github_undated-pv}, which support more functionalities such as fix bug, clean, and customizable prompts. Those tools may already improve the human-AI pair programming interaction in some ways, so future studies could also compare across a variety of LLM-based programming tools.

There is another challenge to describing AI as a pair programmer, following the debate on anthropomorphizing user interfaces \cite{Shneiderman1997-gh} and ongoing discussion as AI demonstrates increasing capabilities to replicate human behaviors \cite{Tan2023-wv, Li2021-rs}. The concern is that anthropomorphized AI could mislead designers and deceive users, impede user agency and responsibility, have deeper ethical and social risks, and may not be more effective anyways. 

However, in educational literature, researchers have been trying to make agents provide naturalistic and human-like interactions with students, using teachable agents \cite{Ogan2012-ip, Bodenheimer2009-sl}, pedagogical agents \cite{Mayer2012-mv, Lin2013-aa, Lusk2007-ga}, conversational agent \cite{Sankaranarayanan2020-fc, Robe2022-rz}, etc. \citet{Kuttal2021-il} explored the trade-off of using a human vs.~AI agent as the pair programming partner. They found that human-human and human-AI led to similar productivity, code quality, and self-efficacy results, and students ``trusted and showed humility towards agents.'' They also found that AI agents successfully facilitated knowledge transfer while failing at providing logical explanations or discussions.

Those anthropomorphized agents mostly seem to be effective in improving learning and motivation \cite{Heidig2011-zw, Schroeder2013-ub}. Some explained the effects using social agency theory \cite{Mayer2012-mv}, cognitive load theory \cite{Lin2013-aa}, and social cues related multimedia learning principles \cite{Mayer2014-wj}.
How well can we apply these theories to LLM-supported AI agents, and what's different in industry versus educational context would be interesting to explore. More works are welcomed to create a shared vocabulary for this field.

\begin{table*}[!h]
\fontsize{8.5}{9}\selectfont
\caption{Challenges in Human-Human Pair Programming \emph{Yield} Opportunities for Human-AI pAIr Programming}
\label{tab:opportunity}
%\vskip 0.1in
\begin{center}
\begin{tabular}{p{0.37\textwidth} | p{0.25\textwidth} | p{0.3\textwidth}}
\toprule
\textbf{Moderating Factors} & \textbf{Human-Human Challenges} & \textbf{Human-AI Opportunities} \\
\midrule
%\abovespace
\textbf{Task Types \& Complexity}: pair work better if the task is not too simple and good for collaboration \cite{Sison2009-mp, Chaparro2005-uk} 
& Hard to design suitable tasks of appropriate complexity level 
& AI may be used to generate collaboration tasks and adjust tasks complexity \\
\midrule%\abovespace
\textbf{Compatibility}: pairs with similar skill levels and compatible working styles work better \cite{Salleh2011-dd, Chaparro2005-uk} & Hard to find a similarly skilled or compatible partner 
& AI partner should adjust to human skill level and adapt to be compatible with different people \\
% Can AI be configured to match human's skill level? \\
\midrule%\abovespace
\textbf{Communication}: pairs work better with productive conversations \cite{Freudenberg2007-jb}, and critiques lead to more debugging success \cite{Murphy2010-mj} 
& Hard to teach effective communication and constructive criticism
& AI partner should support productive conversations and provide critiques \\
% Can AI be configured to support effective interactions like critiques? \\
\midrule%\abovespace
\textbf{Collaboration}: pairs work better with positive interdependence \cite{Preston2006-wi} and clear and balanced responsibilities \cite{Nagappan2003-hb} 
& Hard to teach collaboration and prevent free riders
& AI should support positive social interactions and collaboration and avoid over-assist that eliminates human's need to engage \\
% Can AI be adapted to support the social dynamics of different students? What should we teach students to best collaborate with AI? 
\midrule%\abovespace
\textbf{Logistics}: pair programming is costly to implement because of management challenges \cite{Ally2005-vx, Sun2009-ja} 
& Hard to schedule and assess individual contributions in a pair %resolve conflicts, 
& Scheduling is no longer a problem, but humans should be accountable and responsible when using AI-generated code \\
\bottomrule
\end{tabular}
\end{center}
\end{table*}

\subsection{LLM, A Better pAIr Programmer?}

As reviewed in \cref{mixed-outcome}, previous literature has explored a variety of measures to evaluate different aspects of human-human pair programming, while the current exploration in human-AI pair programming is quite limited. 
\citet{Murillo2023-ld} have proposed evaluation metrics for LLM-based creative code writing assistants in software engineering.
More works could use more valid measures in the human-human pair programming literature to explore how to best help humans and LLM-based AI programming assistant collaborate together. It would also be interesting to have a study setup with three conditions -- human-human, human-AI, and human solo -- working on the same task.

Previous literature suggested some key factors in the success of human-human pair programming, as summarized in \cref{tab:vars}. These moderators that cause challenges for human-human pair programming may yield opportunities to explore in human-AI pair programming (\cref{tab:opportunity}). For example, self-efficacy can lead to a difference in satisfaction \cite{Thomas2003-ud} and gender can lead to a difference in learning \cite{Maguire2014-ee}, do these compatibility moderators influence pAIr too? Can we improve pAIr outcomes using insights derived from human-human literature (e.g., simulate an AI partner with similar self-efficacy levels and the same gender)?
Therefore, in general, we can ask the following questions for future works: \emph{Could these factors be implemented for human-AI pair programming? Would they make human-AI pair programming more effective, less effective, or have no influence, and why?}

\paragraph{Task Types \& Complexity}
As we know from the human-human pair programming literature, a good collaborative task of the right complexity is important, but creating or choosing such tasks can be difficult. % in the industry or educational context
Meanwhile, LLMs help educators efficiently generate instructional materials such as questions \cite{Wang2022-sr}, question-answers \cite{Kumar2022-rb}, feedback \cite{Dai2023-pa}, and hints \cite{Pardos2023-sb}, which could be of similar quality as human-authored content. 
There is also work that suggested the preliminary success in using LLM to break down problems into sub-questions \cite{Sonkar2023-mr}.
Therefore, based on the insight from human-human pair programming literature and the known capacities of LLM, there is an open question to explore in human-AI pAIr programming: can LLM be configured to generate a task type with collaborative learning goals and customize task complexity for a programmer? 

\paragraph{Compatibility - Expertise}
In terms of the compatibility factor expertise, the pair programming literature suggests that matching partners with a similar level of expertise may be the best in promoting productivity and learning \cite{Alves_De_Lima_Salge2016-se, Hannay2009-vv, Chaparro2005-uk}. Evaluation studies show that GPT3-based models can be an above-average student in a CS1 classroom \cite{Puryear2022-db, Finnie-Ansley2022-vm} and its performance gets worse when the code becomes more complicated \cite{Yetistiren2022-so}. GPT4 even does better at solving introductory and basic programming problems (although its correctness is still not comparable to a developer in practice) \cite{Bubeck2023-nd}. We can also purposefully generate bugs and let the models make mistakes \cite{Kang2022-dv}, so potentially, we may create an AI partner with a similar skill level to novice students. Future works can examine how to configure AI to adapt to student's skill levels and whether it will be effective or not.

\paragraph{Other Compatibility Factors}
Researchers have explored how to let LLMs generate interaction based on a designed persona and reasonably replicate human behavior \cite{Aher2022-nw, Horton2023-vo}, and in education, \citet{Cao2023-pp} let LLMs interact with students while role-playing as different fictional characters to help reduce students' anxiety and increase motivation. There are possibilities to personalize an AI partner with different personality traits or the other pair compatibility factors like gender, ethnicity, and self-esteem that \citet{Salleh2011-dd} proposed. Potentially, it can be used to increase programmers' motivation and/or engagement, but how useful it is for human-AI pair programming is yet to be examined. 

\paragraph{Communication}
For communication, we know the social aspect of a conversation matter \cite{Chong2007-uz} and that some types of discourse could be more effective to facilitate debugging \cite{Murphy2010-mj} in human-human pair programming. Therefore, since LLM-based tools such as ChatGPT are able to simulate social interaction, it would be interesting to explore if LLM can support different types of communication, can the different components of communication be replicated in an LLM-based programming assistant, and whether it is effective or not. 

\paragraph{Collaboration}
In terms of collaboration, it is frequently reported that creating smooth collaboration is challenging in both industry \cite{Begel2008-nm} and educational context \cite{Nagappan2003-hb, Williams2002-de}. Given that the free-rider problem reduce pair programming's effectiveness \cite{Nagappan2003-hb} and regular role-switching potentially alleviates the driver's cognitive load and ensures balanced learning outcomes \cite{Alves_De_Lima_Salge2016-se, Umapathy2017-cd}, it would be interesting to explore if LLM-based AI can be configured to avoid over-help, support role-switching, and how to best support the human-AI pair to collaborate.

\paragraph{Logistics}
Logistics-wise, the use of Copilot as a programming partner may have the special advantage of avoiding scheduling logistics, but there are also concerns of accountability that need to be addressed \cite{Finnie-Ansley2022-vm, Bird2023-jv}. In general, there will be ethical risks and social implications of using AI in pair programming at the workplace and in educational contexts, which needs deeper examination in future works.

\subsection{LLM, Students' pAIr Programmer?}

As reviewed in \cref{context}, most current studies that evaluate the efficacy of Copilot are conducted with experienced software developers. If we estimate Copilot's problem-solving abilities as an average student in introductory programming classes, evaluating its performance when pairing up with a professional software developer with much more expertise may not bring enough benefit to the professional. Therefore, working with LLM's current capabilities, it seems like a student-AI pair programming setup would be the most promising to explore, so the next question is: how should we best support student-AI pair programming?

\paragraph{Re-prioritize programming skills.}
Co-working with AI requires a special skill set, and future work could explore how to support students to better develop these crucial skills. \citet{Bird2023-jv} argued that the popularity of LLM-based programming assistants will result in the growing importance of reviewing code as a skill for developers. Nonetheless, in \citet{Perscheid2017-sw}'s interview, none of the professional developers remembered training on debugging at school. There is already rich literature on debugging and testing instructions \cite{Ahrendt2009-sy, Smith2012-ip, McCauley2008-me}, but logistical challenges like the lack of instructional time still exist \cite{McCauley2008-me, Fitzgerald2010-my}, and educators need to better prepare students with debugging and testing skills needed to work with unreliable AI. 

\paragraph{Integrate AIEd frameworks.}
On the theoretical side, 
\citet{Holstein2020-ha} developed a framework to map ways to mutually augment humans and AI in education, for example, by augmenting interpretation, action, scalability, and capacity. Future works can use existing theories in the AI education space to improve the design of the AI pAIr programming partner, and further investigate if LLMs bring new focus and affordances to previous human-AI education frameworks.

\paragraph{Support explanation and communication with students.}
Previous attempts of using AI agent as pair programming partner have shown some preliminary success in knowledge transfer and retention \cite{Robe2022-rz, Han2010-zb}, and the limitation discussed was the lack of discussion and explanation \cite{Kuttal2021-il}. Nowadays, as an LLM-based agent can support more natural interaction and provide good quality explanations in the introductory programming context \cite{Leinonen2023-vz}, it would be interesting to explore if LLM-based AI could resolve some limitations mentioned in pedagogical and conversational agent works before. Self-reflection and explanation techniques may also be adopted to make up for the communication aspect as in human-human pair programming. %\cite{Sarsa2022-ni}

\paragraph{Match expertise with students.}
As discussed in \cref{moderator}, matching expertise is a tricky problem. 
\citet{Lui2006-ur} found that expert-expert pair may not gain as much of an advantage over an expert solo programmer, in comparison to novice-novice pair vs.\ a solo novice. Meanwhile, pairing two novices together raise concerns of ``the blind leading the blind,'' but pairing a novice with an expert may lead to lower self-esteem of the novice \cite{Ally2005-vx}. 
Given all these complexities, when it comes to a student-AI pair and when we only care about the student's learning gains, there are a lot of research questions to ask. If we have full control of the perceived skill level of the AI partner, should we configure it to be similar to the student, slightly higher-skilled, or a lot better? Would it be beneficial to have both a peer AI agent but also a tutor AI agent to assist if students get stuck? 

\paragraph{Avoid over-helping students.}
For programming learners, it would be important to configure the LLM-based programming assistant to avoid over-help. In the few studies that examined novice interaction with Copilot \cite{Prather2023-kj} or a customized programming environment based on LLM-based code generation model Codex \cite{Kazemitabaar2023-gu}. \citet{Prather2023-kj} found that novices do have unique interaction patterns with Copilot and a tendency to rely on and trust the generated code too much. \citet{Kazemitabaar2023-gu} discussed design implications including control over-use and support complete novices. There have also been concerns about academic integrity and changing perception of learning when LLM-based programming tools become easily accessible to students \cite{Becker2023-ue, Puryear2022-db, Prather2023-kj}, which need further explorations for student-AI pair programming.

\paragraph{Boost students' self-confidence.}
Last but not least, pair programming has been shown to benefit students with lower self-efficacy and self-confidence levels \cite{Thomas2003-ud} and women \cite{Maguire2014-ee} more, which could make it a pedagogical tool to engage more vulnerable or underrepresented populations in CS. When an AI is introduced in pair programming, would the same benefit retains? How should we present the AI differently to make it compatible to students with different confidence levels? How do we mitigate the risks of unreliable but seemingly authoritative AI? LLMs may be an opportunity to address some existing challenges in student-student pair programming (as summarized in \cref{tab:opportunity}), but there are still a lot of open questions to ask. %for difficulties such as communication and collaboration training,

\section{Conclusion}

This paper has discussed the concept of human-AI pair programming (pAIr programming). We found that both human-human and human-AI pair programming have  benefits and challenges, but current research did not give us a clear answer on the efficacy of human-AI pair programming. 
Human-human pair programming literature yield insights on what study designs should pAIr researchers adopt (e.g., more realistic observations), what outcomes and measures should pAIr researchers use to evaluate their work (e.g., use more valid quality and productivity measurements, and further investigate cost), and what moderators should pAIr researchers consider to further analyze the pAIr process and improve pAIr design (e.g, compatibility, communication, etc.). 

In conclusion, more valid and comprehensive measurements are needed to evaluate pAIr, more comparisons can be drawn between human-human vs. human-AI pair programming, and more works can explore how to best support LLM-assisted programming with insights from the rich literature on human-human pair programming.

\begin{acks}
Thanks to Ken’s lab members for giving feedback on this work. Thanks to Dr. Stephen MacNeil for coming up with the creative ``pAIr'' keyword for this project.
\end{acks}

\bibliography{paperpile}


\begin{thebibliography}{90}

%%% the Web address.

\ifx \showCODEN    \undefined \def \showCODEN     #1{\unskip}     \fi
\ifx \showDOI      \undefined \def \showDOI       #1{#1}\fi
\ifx \showISBNx    \undefined \def \showISBNx     #1{\unskip}     \fi
\ifx \showISBNxiii \undefined \def \showISBNxiii  #1{\unskip}     \fi
\ifx \showISSN     \undefined \def \showISSN      #1{\unskip}     \fi
\ifx \showLCCN     \undefined \def \showLCCN      #1{\unskip}     \fi
\ifx \shownote     \undefined \def \shownote      #1{#1}          \fi
\ifx \showarticletitle \undefined \def \showarticletitle #1{#1}   \fi
\ifx \showURL      \undefined \def \showURL       {\relax}        \fi
% invisible to TeX
\providecommand\bibfield[2]{#2}
\providecommand\bibinfo[2]{#2}
\providecommand\natexlab[1]{#1}
\providecommand\showeprint[2][]{arXiv:#2}

\bibitem[Aher et~al\mbox{.}(2022)]%
        {Aher2022-nw}
\bibfield{author}{\bibinfo{person}{Gati Aher}, \bibinfo{person}{Rosa~I
  Arriaga}, {and} \bibinfo{person}{Adam~Tauman Kalai}.}
  \bibinfo{year}{2022}\natexlab{}.
\newblock \showarticletitle{Using Large Language Models to Simulate Multiple
  Humans and Replicate Human Subject Studies}.
\newblock  (\bibinfo{date}{Aug.} \bibinfo{year}{2022}).
\newblock
\showeprint[arxiv]{2208.10264}~[cs.CL]
\urldef\tempurl%
\url{http://arxiv.org/abs/2208.10264}
\showURL{%
\tempurl}


\bibitem[Ahrendt et~al\mbox{.}(2009)]%
        {Ahrendt2009-sy}
\bibfield{author}{\bibinfo{person}{Wolfgang Ahrendt}, \bibinfo{person}{Richard
  Bubel}, {and} \bibinfo{person}{Reiner H{\"a}hnle}.}
  \bibinfo{year}{2009}\natexlab{}.
\newblock \showarticletitle{Integrated and {Tool-Supported} Teaching of
  Testing, Debugging, and Verification}. In \bibinfo{booktitle}{\emph{Teaching
  Formal Methods}}. \bibinfo{publisher}{Springer Berlin Heidelberg},
  \bibinfo{pages}{125--143}.
\newblock
\urldef\tempurl%
\url{https://doi.org/10.1007/978-3-642-04912-5\_9}
\showDOI{\tempurl}


\bibitem[Al~Madi(2023)]%
        {Al_Madi2023-fg}
\bibfield{author}{\bibinfo{person}{Naser Al~Madi}.}
  \bibinfo{year}{2023}\natexlab{}.
\newblock \showarticletitle{How Readable is Model-generated Code? Examining
  Readability and Visual Inspection of {GitHub} Copilot}. In
  \bibinfo{booktitle}{\emph{Proceedings of the 37th {IEEE/ACM} International
  Conference on Automated Software Engineering}}.
  \bibinfo{publisher}{Association for Computing Machinery},
  \bibinfo{address}{New York, NY, USA}, \bibinfo{pages}{1--5}.
\newblock
\showISBNx{9781450394758}
\urldef\tempurl%
\url{https://doi.org/10.1145/3551349.3560438}
\showDOI{\tempurl}


\bibitem[Ally et~al\mbox{.}(2005)]%
        {Ally2005-vx}
\bibfield{author}{\bibinfo{person}{Mustafa Ally}, \bibinfo{person}{Fiona
  Darroch}, {and} \bibinfo{person}{Mark Toleman}.}
  \bibinfo{year}{2005}\natexlab{}.
\newblock \showarticletitle{A framework for understanding the factors
  influencing pair programming success}.
\newblock In \bibinfo{booktitle}{\emph{Extreme Programming and Agile Processes
  in Software Engineering}}. \bibinfo{publisher}{Springer Berlin Heidelberg},
  \bibinfo{address}{Berlin, Heidelberg}, \bibinfo{pages}{82--91}.
\newblock
\showISBNx{9783540262770, 9783540314875}
\showISSN{0302-9743, 1611-3349}
\urldef\tempurl%
\url{https://doi.org/10.1007/11499053\_10}
\showDOI{\tempurl}


\bibitem[Alves De Lima~Salge and Berente(2016)]%
        {Alves_De_Lima_Salge2016-se}
\bibfield{author}{\bibinfo{person}{Carolina Alves De Lima~Salge} {and}
  \bibinfo{person}{Nicholas Berente}.} \bibinfo{year}{2016}\natexlab{}.
\newblock \showarticletitle{Pair Programming vs. Solo Programming: What Do We
  Know After 15 Years of Research?}. In \bibinfo{booktitle}{\emph{2016 49th
  Hawaii International Conference on System Sciences ({HICSS})}}.
  \bibinfo{pages}{5398--5406}.
\newblock
\showISSN{1530-1605}
\urldef\tempurl%
\url{https://doi.org/10.1109/HICSS.2016.667}
\showDOI{\tempurl}


\bibitem[Arisholm et~al\mbox{.}(2007)]%
        {Arisholm2007-zk}
\bibfield{author}{\bibinfo{person}{Erik Arisholm}, \bibinfo{person}{Hans
  Gallis}, \bibinfo{person}{Tore Dyba}, {and} \bibinfo{person}{Dag I~K
  Sjoberg}.} \bibinfo{year}{2007}\natexlab{}.
\newblock \showarticletitle{Evaluating Pair Programming with Respect to System
  Complexity and Programmer Expertise}.
\newblock \bibinfo{journal}{\emph{IEEE Trans. Software Eng.}}
  \bibinfo{volume}{33}, \bibinfo{number}{2} (\bibinfo{date}{Feb.}
  \bibinfo{year}{2007}), \bibinfo{pages}{65--86}.
\newblock
\showISSN{1939-3520, 1939-3520}
\urldef\tempurl%
\url{https://doi.org/10.1109/TSE.2007.17}
\showDOI{\tempurl}


\bibitem[Asare et~al\mbox{.}(2022)]%
        {Asare2022-pu}
\bibfield{author}{\bibinfo{person}{Owura Asare}, \bibinfo{person}{Meiyappan
  Nagappan}, {and} \bibinfo{person}{N Asokan}.}
  \bibinfo{year}{2022}\natexlab{}.
\newblock \showarticletitle{Is {GitHub's} Copilot as Bad as Humans at
  Introducing Vulnerabilities in Code?}
\newblock  (\bibinfo{date}{April} \bibinfo{year}{2022}).
\newblock
\showeprint[arxiv]{2204.04741}~[cs.SE]
\urldef\tempurl%
\url{http://arxiv.org/abs/2204.04741}
\showURL{%
\tempurl}


\bibitem[Barke et~al\mbox{.}(2022)]%
        {Barke2022-tv}
\bibfield{author}{\bibinfo{person}{Shraddha Barke}, \bibinfo{person}{Michael~B
  James}, {and} \bibinfo{person}{Nadia Polikarpova}.}
  \bibinfo{year}{2022}\natexlab{}.
\newblock \showarticletitle{Grounded Copilot: How Programmers Interact with
  {Code-Generating} Models}.
\newblock  (\bibinfo{date}{June} \bibinfo{year}{2022}).
\newblock
\showeprint[arxiv]{2206.15000}~[cs.HC]
\urldef\tempurl%
\url{http://arxiv.org/abs/2206.15000}
\showURL{%
\tempurl}


\bibitem[Beck(1999)]%
        {Beck1999-hm}
\bibfield{author}{\bibinfo{person}{Kent Beck}.}
  \bibinfo{year}{1999}\natexlab{}.
\newblock \bibinfo{booktitle}{\emph{Extreme programming explained: embrace
  change}}.
\newblock \bibinfo{publisher}{Addison-Wesley Longman Publishing Co., Inc.},
  \bibinfo{address}{USA}.
\newblock
\showISBNx{9780201616415}
\urldef\tempurl%
\url{https://dl.acm.org/doi/10.5555/318762}
\showURL{%
\tempurl}


\bibitem[Becker et~al\mbox{.}(2023)]%
        {Becker2023-ue}
\bibfield{author}{\bibinfo{person}{Brett~A Becker}, \bibinfo{person}{Paul
  Denny}, \bibinfo{person}{James Finnie-Ansley}, \bibinfo{person}{Andrew
  Luxton-Reilly}, \bibinfo{person}{James Prather}, {and}
  \bibinfo{person}{Eddie~Antonio Santos}.} \bibinfo{year}{2023}\natexlab{}.
\newblock \showarticletitle{Programming Is Hard - Or at Least It Used to Be:
  Educational Opportunities and Challenges of {AI} Code Generation}. In
  \bibinfo{booktitle}{\emph{Proceedings of the 54th {ACM} Technical Symposium
  on Computer Science Education V. 1}} (Toronto ON, Canada)
  \emph{(\bibinfo{series}{SIGCSE 2023})}. \bibinfo{publisher}{Association for
  Computing Machinery}, \bibinfo{address}{New York, NY, USA},
  \bibinfo{pages}{500--506}.
\newblock
\showISBNx{9781450394314}
\urldef\tempurl%
\url{https://doi.org/10.1145/3545945.3569759}
\showDOI{\tempurl}


\bibitem[Begel and Nagappan(2008)]%
        {Begel2008-nm}
\bibfield{author}{\bibinfo{person}{Andrew Begel} {and}
  \bibinfo{person}{Nachiappan Nagappan}.} \bibinfo{year}{2008}\natexlab{}.
\newblock \showarticletitle{Pair programming: what's in it for me?}. In
  \bibinfo{booktitle}{\emph{Proceedings of the Second {ACM-IEEE} international
  symposium on Empirical software engineering and measurement}} (Kaiserslautern
  Germany). \bibinfo{publisher}{ACM}, \bibinfo{address}{New York, NY, USA}.
\newblock
\showISBNx{9781595939715}
\urldef\tempurl%
\url{https://doi.org/10.1145/1414004.1414026}
\showDOI{\tempurl}


\bibitem[Bird et~al\mbox{.}(2023)]%
        {Bird2023-jv}
\bibfield{author}{\bibinfo{person}{Christian Bird}, \bibinfo{person}{Denae
  Ford}, \bibinfo{person}{Thomas Zimmermann}, \bibinfo{person}{Nicole
  Forsgren}, \bibinfo{person}{Eirini Kalliamvakou}, \bibinfo{person}{Travis
  Lowdermilk}, {and} \bibinfo{person}{Idan Gazit}.}
  \bibinfo{year}{2023}\natexlab{}.
\newblock \showarticletitle{Taking Flight with Copilot: Early insights and
  opportunities of {AI-powered} pair-programming tools}.
\newblock \bibinfo{journal}{\emph{Queueing Syst.}} \bibinfo{volume}{20},
  \bibinfo{number}{6} (\bibinfo{date}{Jan.} \bibinfo{year}{2023}),
  \bibinfo{pages}{35--57}.
\newblock
\showISSN{0257-0130, 1542-7730}
\urldef\tempurl%
\url{https://doi.org/10.1145/3582083}
\showDOI{\tempurl}


\bibitem[Bodenheimer et~al\mbox{.}(2009)]%
        {Bodenheimer2009-sl}
\bibfield{author}{\bibinfo{person}{Bobby Bodenheimer}, \bibinfo{person}{B
  Sanders}, \bibinfo{person}{M~R Kramer}, \bibinfo{person}{K Viswanath},
  \bibinfo{person}{R Balachandran}, \bibinfo{person}{Kadira Belynne}, {and}
  \bibinfo{person}{Gautam Biswas}.} \bibinfo{year}{2009}\natexlab{}.
\newblock \showarticletitle{Construction and Evaluation of Animated Teachable
  Agents}.
\newblock \bibinfo{journal}{\emph{J. Educ. Technol. Soc.}}
  (\bibinfo{year}{2009}).
\newblock
\urldef\tempurl%
\url{https://www.semanticscholar.org/paper/2899ac4dfe209db4767ec01b5df337079bada517}
\showURL{%
\tempurl}


\bibitem[Bubeck et~al\mbox{.}(2023)]%
        {Bubeck2023-nd}
\bibfield{author}{\bibinfo{person}{S{\'e}bastien Bubeck},
  \bibinfo{person}{Varun Chandrasekaran}, \bibinfo{person}{Ronen Eldan},
  \bibinfo{person}{Johannes Gehrke}, \bibinfo{person}{Eric Horvitz},
  \bibinfo{person}{Ece Kamar}, \bibinfo{person}{Peter Lee},
  \bibinfo{person}{Yin~Tat Lee}, \bibinfo{person}{Yuanzhi Li},
  \bibinfo{person}{Scott Lundberg}, \bibinfo{person}{Harsha Nori},
  \bibinfo{person}{Hamid Palangi}, \bibinfo{person}{Marco~Tulio Ribeiro}, {and}
  \bibinfo{person}{Yi Zhang}.} \bibinfo{year}{2023}\natexlab{}.
\newblock \showarticletitle{Sparks of Artificial General Intelligence: Early
  experiments with {GPT-4}}.
\newblock  (\bibinfo{date}{March} \bibinfo{year}{2023}).
\newblock
\showeprint[arxiv]{2303.12712}~[cs.CL]
\urldef\tempurl%
\url{http://arxiv.org/abs/2303.12712}
\showURL{%
\tempurl}


\bibitem[Cao(2023)]%
        {Cao2023-pp}
\bibfield{author}{\bibinfo{person}{Chen Cao}.} \bibinfo{year}{2023}\natexlab{}.
\newblock \showarticletitle{Scaffolding {CS1} Courses with a Large Language
  {Model-Powered} Intelligent Tutoring System}. In
  \bibinfo{booktitle}{\emph{Companion Proceedings of the 28th International
  Conference on Intelligent User Interfaces}} (Sydney, NSW, Australia)
  \emph{(\bibinfo{series}{IUI '23 Companion})}. \bibinfo{publisher}{Association
  for Computing Machinery}, \bibinfo{address}{New York, NY, USA},
  \bibinfo{pages}{229--232}.
\newblock
\showISBNx{9798400701078}
\urldef\tempurl%
\url{https://doi.org/10.1145/3581754.3584111}
\showDOI{\tempurl}


\bibitem[Chaparro et~al\mbox{.}(2005)]%
        {Chaparro2005-uk}
\bibfield{author}{\bibinfo{person}{E~A Chaparro}, \bibinfo{person}{Aybala
  Yuksel}, \bibinfo{person}{Pablo Romero}, {and} \bibinfo{person}{Sallyann
  Bryant}.} \bibinfo{year}{2005}\natexlab{}.
\newblock \showarticletitle{Factors Affecting the Perceived Effectiveness of
  Pair Programming in Higher Education}.
\newblock \bibinfo{journal}{\emph{Annual Workshop of the Psychology of
  Programming Interest Group}} (\bibinfo{year}{2005}).
\newblock
\urldef\tempurl%
\url{https://www.semanticscholar.org/paper/c095f0d9b17cd9c2851000534740e7cc087253fa}
\showURL{%
\tempurl}


\bibitem[Chong and Hurlbutt(2007)]%
        {Chong2007-uz}
\bibfield{author}{\bibinfo{person}{Jan Chong} {and} \bibinfo{person}{Tom
  Hurlbutt}.} \bibinfo{year}{2007}\natexlab{}.
\newblock \showarticletitle{The Social Dynamics of Pair Programming}. In
  \bibinfo{booktitle}{\emph{29th International Conference on Software
  Engineering ({ICSE'07})}}. \bibinfo{publisher}{ieeexplore.ieee.org},
  \bibinfo{pages}{354--363}.
\newblock
\showISSN{1558-1225}
\urldef\tempurl%
\url{https://doi.org/10.1109/ICSE.2007.87}
\showDOI{\tempurl}


\bibitem[Cockburn and Williams(2001)]%
        {Cockburn2001-ci}
\bibfield{author}{\bibinfo{person}{Alistair Cockburn} {and} \bibinfo{person}{L
  Williams}.} \bibinfo{year}{2001}\natexlab{}.
\newblock \showarticletitle{The costs and benefits of pair programming}.
\newblock \bibinfo{journal}{\emph{Computer Science}} (\bibinfo{year}{2001}).
\newblock
\urldef\tempurl%
\url{https://www.semanticscholar.org/paper/5ff7b75b20fdbfae23587b660b7093aec2f48e69}
\showURL{%
\tempurl}


\bibitem[da~Silva~Est{\'a}cio and Prikladnicki(2015)]%
        {Da_Silva_Estacio2015-nh}
\bibfield{author}{\bibinfo{person}{Bernardo~Jos{\'e} da Silva~Est{\'a}cio}
  {and} \bibinfo{person}{Rafael Prikladnicki}.}
  \bibinfo{year}{2015}\natexlab{}.
\newblock \showarticletitle{Distributed Pair Programming: A Systematic
  Literature Review}.
\newblock \bibinfo{journal}{\emph{Information and Software Technology}}
  \bibinfo{volume}{63} (\bibinfo{date}{July} \bibinfo{year}{2015}),
  \bibinfo{pages}{1--10}.
\newblock
\showISSN{0950-5849}
\urldef\tempurl%
\url{https://doi.org/10.1016/j.infsof.2015.02.011}
\showDOI{\tempurl}


\bibitem[Dai et~al\mbox{.}(2023)]%
        {Dai2023-pa}
\bibfield{author}{\bibinfo{person}{Wei Dai}, \bibinfo{person}{Jionghao Lin},
  \bibinfo{person}{Flora Jin}, \bibinfo{person}{Tongguang Li},
  \bibinfo{person}{Yi-Shan Tsai}, \bibinfo{person}{Dragan Gasevic}, {and}
  \bibinfo{person}{Guanliang Chen}.} \bibinfo{year}{2023}\natexlab{}.
\newblock \bibinfo{title}{Can Large Language Models Provide Feedback to
  Students? A Case Study on {ChatGPT}}.  (\bibinfo{date}{April}
  \bibinfo{year}{2023}).
\newblock
\urldef\tempurl%
\url{https://doi.org/10.35542/osf.io/hcgzj}
\showDOI{\tempurl}


\bibitem[Dakhel et~al\mbox{.}(2022)]%
        {Dakhel2022-he}
\bibfield{author}{\bibinfo{person}{Arghavan~Moradi Dakhel},
  \bibinfo{person}{Vahid Majdinasab}, \bibinfo{person}{Amin Nikanjam},
  \bibinfo{person}{Foutse Khomh}, \bibinfo{person}{Michel~C Desmarais},
  \bibinfo{person}{Zhen Ming}, {and} \bibinfo{person}{{Jiang}}.}
  \bibinfo{year}{2022}\natexlab{}.
\newblock \showarticletitle{{GitHub} Copilot {AI} pair programmer: Asset or
  Liability?}
\newblock \bibinfo{journal}{\emph{ArXiv}} (\bibinfo{year}{2022}).
\newblock
\urldef\tempurl%
\url{https://doi.org/10.48550/ARXIV.2206.15331}
\showDOI{\tempurl}


\bibitem[Finnie-Ansley et~al\mbox{.}(2022)]%
        {Finnie-Ansley2022-vm}
\bibfield{author}{\bibinfo{person}{James Finnie-Ansley}, \bibinfo{person}{Paul
  Denny}, \bibinfo{person}{Brett~A Becker}, \bibinfo{person}{Andrew
  Luxton-Reilly}, {and} \bibinfo{person}{James Prather}.}
  \bibinfo{year}{2022}\natexlab{}.
\newblock \showarticletitle{The Robots Are Coming: Exploring the Implications
  of {OpenAI} Codex on Introductory Programming}. In
  \bibinfo{booktitle}{\emph{Australasian Computing Education Conference}}
  (Virtual Event, Australia) \emph{(\bibinfo{series}{ACE '22})}.
  \bibinfo{publisher}{Association for Computing Machinery},
  \bibinfo{address}{New York, NY, USA}, \bibinfo{pages}{10--19}.
\newblock
\showISBNx{9781450396431}
\urldef\tempurl%
\url{https://doi.org/10.1145/3511861.3511863}
\showDOI{\tempurl}


\bibitem[Fitzgerald et~al\mbox{.}(2010)]%
        {Fitzgerald2010-my}
\bibfield{author}{\bibinfo{person}{Sue Fitzgerald}, \bibinfo{person}{Ren{\'e}e
  McCauley}, \bibinfo{person}{Brian Hanks}, \bibinfo{person}{Laurie Murphy},
  \bibinfo{person}{Beth Simon}, {and} \bibinfo{person}{Carol Zander}.}
  \bibinfo{year}{2010}\natexlab{}.
\newblock \showarticletitle{Debugging From the Student Perspective}.
\newblock \bibinfo{journal}{\emph{IEEE Trans. Educ.}} \bibinfo{volume}{53},
  \bibinfo{number}{3} (\bibinfo{date}{Aug.} \bibinfo{year}{2010}),
  \bibinfo{pages}{390--396}.
\newblock
\showISSN{1557-9638}
\urldef\tempurl%
\url{https://doi.org/10.1109/TE.2009.2025266}
\showDOI{\tempurl}


\bibitem[Freudenberg et~al\mbox{.}(2007)]%
        {Freudenberg2007-jb}
\bibfield{author}{\bibinfo{person}{S Freudenberg}, \bibinfo{person}{Pablo
  Romero}, {and} \bibinfo{person}{Benedict Du~Boulay}.}
  \bibinfo{year}{2007}\natexlab{}.
\newblock \showarticletitle{Talking the talk: Is intermediate-level
  conversation the key to the pair programming success story?}. In
  \bibinfo{booktitle}{\emph{{AGILE} 2007}}. \bibinfo{publisher}{unknown},
  \bibinfo{pages}{84--91}.
\newblock
\urldef\tempurl%
\url{https://doi.org/10.1109/AGILE.2007.1}
\showDOI{\tempurl}


\bibitem[{Github}({[n.\,d.]})]%
        {Github_undated-pv}
\bibfield{author}{\bibinfo{person}{{Github}}.}
  \bibinfo{year}{[n.\,d.]}\natexlab{}.
\newblock \bibinfo{title}{{GitHub} Copilot Labs}.
\newblock
  \bibinfo{howpublished}{\url{https://githubnext.com/projects/copilot-labs/}}.
\newblock
\urldef\tempurl%
\url{https://githubnext.com/projects/copilot-labs/}
\showURL{%
\tempurl}
\newblock
\shownote{Accessed: 2023-5-19}.


\bibitem[{GitHub}(2021)]%
        {GitHub2021-kz}
\bibfield{author}{\bibinfo{person}{{GitHub}}.} \bibinfo{year}{2021}\natexlab{}.
\newblock \bibinfo{title}{Your {AI} pair programmer: Copilot}.
\newblock \bibinfo{howpublished}{\url{https://github.com/features/copilot}}.
\newblock
\urldef\tempurl%
\url{https://github.com/features/copilot}
\showURL{%
\tempurl}
\newblock
\shownote{Accessed: 2022-10-5}.


\bibitem[{Google}({[n.\,d.]})]%
        {Google_undated-lx}
\bibfield{author}{\bibinfo{person}{{Google}}.}
  \bibinfo{year}{[n.\,d.]}\natexlab{}.
\newblock \bibinfo{title}{Bard}.
\newblock \bibinfo{howpublished}{\url{https://bard.google.com/}}.
\newblock
\urldef\tempurl%
\url{https://bard.google.com/}
\showURL{%
\tempurl}
\newblock
\shownote{Accessed: 2023-5-19}.


\bibitem[Han et~al\mbox{.}(2010)]%
        {Han2010-zb}
\bibfield{author}{\bibinfo{person}{Keun-Woo Han}, \bibinfo{person}{Eunkyoung
  Lee}, {and} \bibinfo{person}{Youngjun Lee}.} \bibinfo{year}{2010}\natexlab{}.
\newblock \showarticletitle{The Impact of a {Peer-Learning} Agent Based on Pair
  Programming in a Programming Course}.
\newblock \bibinfo{journal}{\emph{IEEE Trans. Educ.}} \bibinfo{volume}{53},
  \bibinfo{number}{2} (\bibinfo{date}{May} \bibinfo{year}{2010}),
  \bibinfo{pages}{318--327}.
\newblock
\showISSN{1557-9638}
\urldef\tempurl%
\url{https://doi.org/10.1109/TE.2009.2019121}
\showDOI{\tempurl}


\bibitem[Hanks et~al\mbox{.}(2011)]%
        {Hanks2011-ae}
\bibfield{author}{\bibinfo{person}{Brian Hanks}, \bibinfo{person}{Sue
  Fitzgerald}, \bibinfo{person}{Ren{\'e}e McCauley}, \bibinfo{person}{Laurie
  Murphy}, {and} \bibinfo{person}{Carol Zander}.}
  \bibinfo{year}{2011}\natexlab{}.
\newblock \showarticletitle{Pair programming in education: a literature
  review}.
\newblock \bibinfo{journal}{\emph{Comput. Sci. Educ.}} \bibinfo{volume}{21},
  \bibinfo{number}{2} (\bibinfo{date}{June} \bibinfo{year}{2011}),
  \bibinfo{pages}{135--173}.
\newblock
\showISSN{0899-3408, 1744-5175}
\urldef\tempurl%
\url{https://doi.org/10.1080/08993408.2011.579808}
\showDOI{\tempurl}


\bibitem[Hannay et~al\mbox{.}(2010)]%
        {Hannay2010-dc}
\bibfield{author}{\bibinfo{person}{Jo~E Hannay}, \bibinfo{person}{Erik
  Arisholm}, \bibinfo{person}{Harald Engvik}, {and} \bibinfo{person}{Dag I~K
  Sjoberg}.} \bibinfo{year}{2010}\natexlab{}.
\newblock \showarticletitle{Effects of Personality on Pair Programming}.
\newblock \bibinfo{journal}{\emph{IEEE Trans. Software Eng.}}
  \bibinfo{volume}{36}, \bibinfo{number}{1} (\bibinfo{date}{Jan.}
  \bibinfo{year}{2010}), \bibinfo{pages}{61--80}.
\newblock
\showISSN{1939-3520}
\urldef\tempurl%
\url{https://doi.org/10.1109/TSE.2009.41}
\showDOI{\tempurl}


\bibitem[Hannay et~al\mbox{.}(2009)]%
        {Hannay2009-vv}
\bibfield{author}{\bibinfo{person}{Jo~E Hannay}, \bibinfo{person}{Tore
  Dyb{\aa}}, \bibinfo{person}{Erik Arisholm}, {and} \bibinfo{person}{Dag I~K
  Sj{\o}berg}.} \bibinfo{year}{2009}\natexlab{}.
\newblock \showarticletitle{The effectiveness of pair programming: A
  meta-analysis}.
\newblock \bibinfo{journal}{\emph{Information and Software Technology}}
  \bibinfo{volume}{51}, \bibinfo{number}{7} (\bibinfo{date}{July}
  \bibinfo{year}{2009}), \bibinfo{pages}{1110--1122}.
\newblock
\showISSN{0950-5849}
\urldef\tempurl%
\url{https://doi.org/10.1016/j.infsof.2009.02.001}
\showDOI{\tempurl}


\bibitem[Heidig and Clarebout(2011)]%
        {Heidig2011-zw}
\bibfield{author}{\bibinfo{person}{Steffi Heidig} {and}
  \bibinfo{person}{Geraldine Clarebout}.} \bibinfo{year}{2011}\natexlab{}.
\newblock \showarticletitle{Do pedagogical agents make a difference to student
  motivation and learning?}
\newblock \bibinfo{journal}{\emph{Educational Research Review}}
  \bibinfo{volume}{6}, \bibinfo{number}{1} (\bibinfo{date}{Jan.}
  \bibinfo{year}{2011}), \bibinfo{pages}{27--54}.
\newblock
\showISSN{1747-938X}
\urldef\tempurl%
\url{https://doi.org/10.1016/j.edurev.2010.07.004}
\showDOI{\tempurl}


\bibitem[Holstein et~al\mbox{.}(2020)]%
        {Holstein2020-ha}
\bibfield{author}{\bibinfo{person}{Kenneth Holstein}, \bibinfo{person}{Vincent
  Aleven}, {and} \bibinfo{person}{Nikol Rummel}.}
  \bibinfo{year}{2020}\natexlab{}.
\newblock \showarticletitle{A Conceptual Framework for {Human--AI} Hybrid
  Adaptivity in Education}.
\newblock \bibinfo{journal}{\emph{Artificial Intelligence in Education}}
  \bibinfo{volume}{12163} (\bibinfo{date}{June} \bibinfo{year}{2020}),
  \bibinfo{pages}{240}.
\newblock
\urldef\tempurl%
\url{https://doi.org/10.1007/978-3-030-52237-7\_20}
\showDOI{\tempurl}


\bibitem[Horton(2023)]%
        {Horton2023-vo}
\bibfield{author}{\bibinfo{person}{John~J Horton}.}
  \bibinfo{year}{2023}\natexlab{}.
\newblock \showarticletitle{Large Language Models as Simulated Economic Agents:
  What Can We Learn from Homo Silicus?}
\newblock  (\bibinfo{date}{Jan.} \bibinfo{year}{2023}).
\newblock
\showeprint[arxiv]{2301.07543}~[econ.GN]
\urldef\tempurl%
\url{http://arxiv.org/abs/2301.07543}
\showURL{%
\tempurl}


\bibitem[Imai(2022)]%
        {Imai2022-so}
\bibfield{author}{\bibinfo{person}{Saki Imai}.}
  \bibinfo{year}{2022}\natexlab{}.
\newblock \showarticletitle{Is {GitHub} Copilot a Substitute for Human
  Pair-programming? An Empirical Study}. In \bibinfo{booktitle}{\emph{2022
  {IEEE/ACM} 44th International Conference on Software Engineering: Companion
  Proceedings ({ICSE-Companion})}}. \bibinfo{publisher}{ieeexplore.ieee.org},
  \bibinfo{pages}{319--321}.
\newblock
\showISSN{2574-1926}
\urldef\tempurl%
\url{https://doi.org/10.1145/3510454.3522684}
\showDOI{\tempurl}


\bibitem[Jensen(2005)]%
        {Jensen2005-jb}
\bibfield{author}{\bibinfo{person}{Randall~W Jensen}.}
  \bibinfo{year}{2005}\natexlab{}.
\newblock \showarticletitle{A Pair Programming Experience}.
\newblock \bibinfo{journal}{\emph{ACCU - professionalism in programming
  Overload}} \bibinfo{volume}{13}, \bibinfo{number}{65} (\bibinfo{date}{Feb.}
  \bibinfo{year}{2005}).
\newblock
\urldef\tempurl%
\url{https://accu.org/journals/overload/13/65/jensen_254/}
\showURL{%
\tempurl}


\bibitem[Kalliamvakou(2022)]%
        {Kalliamvakou2022-ku}
\bibfield{author}{\bibinfo{person}{Eirini Kalliamvakou}.}
  \bibinfo{year}{2022}\natexlab{}.
\newblock \bibinfo{title}{Research: quantifying {GitHub} Copilot's impact on
  developer productivity and happiness}.
\newblock
  \bibinfo{howpublished}{\url{https://github.blog/2022-09-07-research-quantifying-github-copilots-impact-on-developer-productivity-and-happiness/}}.
\newblock
\urldef\tempurl%
\url{https://github.blog/2022-09-07-research-quantifying-github-copilots-impact-on-developer-productivity-and-happiness/}
\showURL{%
\tempurl}
\newblock
\shownote{Accessed: 2022-10-13}.


\bibitem[Kang et~al\mbox{.}(2022)]%
        {Kang2022-dv}
\bibfield{author}{\bibinfo{person}{Sungmin Kang}, \bibinfo{person}{Juyeon
  Yoon}, {and} \bibinfo{person}{Shin Yoo}.} \bibinfo{year}{2022}\natexlab{}.
\newblock \showarticletitle{Large Language Models are few-shot testers:
  Exploring {LLM-based} general bug reproduction}.
\newblock \bibinfo{journal}{\emph{ArXiv}} (\bibinfo{year}{2022}).
\newblock
\urldef\tempurl%
\url{https://doi.org/10.48550/ARXIV.2209.11515}
\showDOI{\tempurl}


\bibitem[Kazemitabaar et~al\mbox{.}(2023)]%
        {Kazemitabaar2023-gu}
\bibfield{author}{\bibinfo{person}{Majeed Kazemitabaar},
  \bibinfo{person}{Justin Chow}, \bibinfo{person}{Carl Ka~To Ma},
  \bibinfo{person}{Barbara~J Ericson}, \bibinfo{person}{David Weintrop}, {and}
  \bibinfo{person}{Tovi Grossman}.} \bibinfo{year}{2023}\natexlab{}.
\newblock \showarticletitle{Studying the effect of {AI} Code Generators on
  Supporting Novice Learners in Introductory Programming}.
\newblock  (\bibinfo{date}{Feb.} \bibinfo{year}{2023}).
\newblock
\showeprint[arxiv]{2302.07427}~[cs.HC]
\urldef\tempurl%
\url{http://arxiv.org/abs/2302.07427}
\showURL{%
\tempurl}


\bibitem[Kumar et~al\mbox{.}(2022)]%
        {Kumar2022-rb}
\bibfield{author}{\bibinfo{person}{Shobhan Kumar}, \bibinfo{person}{Arun
  Chauhan}, {and} \bibinfo{person}{Pavan Kumar~C.}}
  \bibinfo{year}{2022}\natexlab{}.
\newblock \showarticletitle{Learning Enhancement Using {Question-Answer}
  Generation for e-Book Using Contrastive {Fine-Tuned} {T5}}. In
  \bibinfo{booktitle}{\emph{Big Data Analytics}}. \bibinfo{publisher}{Springer
  Nature Switzerland}, \bibinfo{pages}{68--87}.
\newblock
\urldef\tempurl%
\url{https://doi.org/10.1007/978-3-031-24094-2\_5}
\showDOI{\tempurl}


\bibitem[Kuttal et~al\mbox{.}(2021)]%
        {Kuttal2021-il}
\bibfield{author}{\bibinfo{person}{Sandeep~Kaur Kuttal}, \bibinfo{person}{Bali
  Ong}, \bibinfo{person}{Kate Kwasny}, {and} \bibinfo{person}{Peter Robe}.}
  \bibinfo{year}{2021}\natexlab{}.
\newblock \showarticletitle{Trade-offs for Substituting a Human with an Agent
  in a Pair Programming Context: The Good, the Bad, and the Ugly}. In
  \bibinfo{booktitle}{\emph{Proceedings of the 2021 {CHI} Conference on Human
  Factors in Computing Systems}} (Yokohama, Japan) \emph{(\bibinfo{series}{CHI
  '21}, \bibinfo{number}{Article 243})}. \bibinfo{publisher}{Association for
  Computing Machinery}, \bibinfo{address}{New York, NY, USA},
  \bibinfo{pages}{1--20}.
\newblock
\showISBNx{9781450380966}
\urldef\tempurl%
\url{https://doi.org/10.1145/3411764.3445659}
\showDOI{\tempurl}


\bibitem[Leinonen et~al\mbox{.}(2023)]%
        {Leinonen2023-vz}
\bibfield{author}{\bibinfo{person}{Juho Leinonen}, \bibinfo{person}{Paul
  Denny}, \bibinfo{person}{Stephen MacNeil}, \bibinfo{person}{Sami Sarsa},
  \bibinfo{person}{Seth Bernstein}, \bibinfo{person}{Joanne Kim},
  \bibinfo{person}{Andrew Tran}, {and} \bibinfo{person}{Arto Hellas}.}
  \bibinfo{year}{2023}\natexlab{}.
\newblock \showarticletitle{Comparing Code Explanations Created by Students and
  Large Language Models}.
\newblock  (\bibinfo{date}{April} \bibinfo{year}{2023}).
\newblock
\showeprint[arxiv]{2304.03938}~[cs.CY]
\urldef\tempurl%
\url{http://arxiv.org/abs/2304.03938}
\showURL{%
\tempurl}


\bibitem[Li and Suh(2021)]%
        {Li2021-rs}
\bibfield{author}{\bibinfo{person}{Mengjun Li} {and} \bibinfo{person}{Ayoung
  Suh}.} \bibinfo{year}{2021}\natexlab{}.
\newblock \showarticletitle{Machinelike or Humanlike? A Literature Review of
  Anthropomorphism in {AI-Enabled} Technology}. In
  \bibinfo{booktitle}{\emph{Hawaii International Conference on System Sciences
  2021 ({HICSS-54})}}.
\newblock
\urldef\tempurl%
\url{https://aisel.aisnet.org/hicss-54/in/ai_based_assistants/5/}
\showURL{%
\tempurl}


\bibitem[Lin et~al\mbox{.}(2013)]%
        {Lin2013-aa}
\bibfield{author}{\bibinfo{person}{Lijia Lin}, \bibinfo{person}{Robert~K
  Atkinson}, \bibinfo{person}{Robert~M Christopherson},
  \bibinfo{person}{Stacey~S Joseph}, {and} \bibinfo{person}{Caroline~J
  Harrison}.} \bibinfo{year}{2013}\natexlab{}.
\newblock \showarticletitle{Animated agents and learning: Does the type of
  verbal feedback they provide matter?}
\newblock \bibinfo{journal}{\emph{Comput. Educ.}}  \bibinfo{volume}{67}
  (\bibinfo{date}{Sept.} \bibinfo{year}{2013}), \bibinfo{pages}{239--249}.
\newblock
\showISSN{0360-1315}
\urldef\tempurl%
\url{https://doi.org/10.1016/j.compedu.2013.04.017}
\showDOI{\tempurl}


\bibitem[Lui and Chan(2006)]%
        {Lui2006-ur}
\bibfield{author}{\bibinfo{person}{Kim~Man Lui} {and} \bibinfo{person}{Keith
  C~C Chan}.} \bibinfo{year}{2006}\natexlab{}.
\newblock \showarticletitle{Pair programming productivity: Novice--novice vs.
  expert--expert}.
\newblock \bibinfo{journal}{\emph{Int. J. Hum. Comput. Stud.}}
  \bibinfo{volume}{64}, \bibinfo{number}{9} (\bibinfo{date}{Sept.}
  \bibinfo{year}{2006}), \bibinfo{pages}{915--925}.
\newblock
\showISSN{1071-5819, 1095-9300}
\urldef\tempurl%
\url{https://doi.org/10.1016/j.ijhcs.2006.04.010}
\showDOI{\tempurl}


\bibitem[Lusk and Atkinson(2007)]%
        {Lusk2007-ga}
\bibfield{author}{\bibinfo{person}{Mary~Margaret Lusk} {and}
  \bibinfo{person}{Robert~K Atkinson}.} \bibinfo{year}{2007}\natexlab{}.
\newblock \showarticletitle{Animated pedagogical agents: does their degree of
  embodiment impact learning from static or animated worked examples?}
\newblock \bibinfo{journal}{\emph{Appl. Cogn. Psychol.}} \bibinfo{volume}{21},
  \bibinfo{number}{6} (\bibinfo{date}{Sept.} \bibinfo{year}{2007}),
  \bibinfo{pages}{747--764}.
\newblock
\showISSN{0888-4080, 1099-0720}
\urldef\tempurl%
\url{https://doi.org/10.1002/acp.1347}
\showDOI{\tempurl}


\bibitem[Maguire et~al\mbox{.}(2014)]%
        {Maguire2014-ee}
\bibfield{author}{\bibinfo{person}{Phil Maguire}, \bibinfo{person}{Rebecca
  Maguire}, \bibinfo{person}{Philip Hyland}, {and} \bibinfo{person}{Patrick
  Marshall}.} \bibinfo{year}{2014}\natexlab{}.
\newblock \showarticletitle{Enhancing collaborative learning using pair
  programming: Who benefits?}
\newblock \bibinfo{journal}{\emph{AISHE-J}} \bibinfo{volume}{6},
  \bibinfo{number}{2} (\bibinfo{date}{June} \bibinfo{year}{2014}).
\newblock
\showISSN{2009-3160, 2009-3160}
\urldef\tempurl%
\url{https://ojs.aishe.org/index.php/aishe-j/article/view/141}
\showURL{%
\tempurl}


\bibitem[Mayer(2014)]%
        {Mayer2014-wj}
\bibfield{author}{\bibinfo{person}{Richard~E Mayer}.}
  \bibinfo{year}{2014}\natexlab{}.
\newblock \showarticletitle{Principles based on social cues in multimedia
  learning: Personalization, voice, image, and embodiment principles}.
\newblock \bibinfo{journal}{\emph{The Cambridge handbook of multimedia
  learning}}  \bibinfo{volume}{16} (\bibinfo{year}{2014}),
  \bibinfo{pages}{345--370}.
\newblock
\urldef\tempurl%
\url{https://books.google.com/books?hl=en&lr=&id=r3rsAwAAQBAJ&oi=fnd&pg=PA345&ots=iUhQ53T8QY&sig=5tQyKi_f-7aILMxRLuwGTLIix3c}
\showURL{%
\tempurl}


\bibitem[Mayer and DaPra(2012)]%
        {Mayer2012-mv}
\bibfield{author}{\bibinfo{person}{Richard~E Mayer} {and}
  \bibinfo{person}{C~Scott DaPra}.} \bibinfo{year}{2012}\natexlab{}.
\newblock \showarticletitle{An embodiment effect in computer-based learning
  with animated pedagogical agents}.
\newblock \bibinfo{journal}{\emph{J. Exp. Psychol. Appl.}}
  \bibinfo{volume}{18}, \bibinfo{number}{3} (\bibinfo{date}{Sept.}
  \bibinfo{year}{2012}), \bibinfo{pages}{239--252}.
\newblock
\showISSN{1076-898X, 1939-2192}
\urldef\tempurl%
\url{https://doi.org/10.1037/a0028616}
\showDOI{\tempurl}


\bibitem[McCauley et~al\mbox{.}(2008)]%
        {McCauley2008-me}
\bibfield{author}{\bibinfo{person}{Renee McCauley}, \bibinfo{person}{Sue
  Fitzgerald}, \bibinfo{person}{Gary Lewandowski}, \bibinfo{person}{Laurie
  Murphy}, \bibinfo{person}{Beth Simon}, \bibinfo{person}{Lynda Thomas}, {and}
  \bibinfo{person}{Carol Zander}.} \bibinfo{year}{2008}\natexlab{}.
\newblock \showarticletitle{Debugging: A Review of the Literature from an
  Educational Perspective}.
\newblock \bibinfo{journal}{\emph{Computer Science Education}}
  \bibinfo{volume}{18}, \bibinfo{number}{2} (\bibinfo{date}{June}
  \bibinfo{year}{2008}), \bibinfo{pages}{67--92}.
\newblock
\showISSN{0899-3408}
\urldef\tempurl%
\url{https://doi.org/10.1080/08993400802114581}
\showDOI{\tempurl}


\bibitem[McDowell et~al\mbox{.}(2002)]%
        {McDowell2002-bi}
\bibfield{author}{\bibinfo{person}{Charlie McDowell}, \bibinfo{person}{Linda
  Werner}, \bibinfo{person}{Heather Bullock}, {and} \bibinfo{person}{Julian
  Fernald}.} \bibinfo{year}{2002}\natexlab{}.
\newblock \showarticletitle{The effects of pair-programming on performance in
  an introductory programming course}. In \bibinfo{booktitle}{\emph{Proceedings
  of the 33rd {SIGCSE} technical symposium on Computer science education}}
  (Cincinnati, Kentucky) \emph{(\bibinfo{series}{SIGCSE '02})}.
  \bibinfo{publisher}{Association for Computing Machinery},
  \bibinfo{address}{New York, NY, USA}, \bibinfo{pages}{38--42}.
\newblock
\showISBNx{9781581134735}
\urldef\tempurl%
\url{https://doi.org/10.1145/563340.563353}
\showDOI{\tempurl}


\bibitem[McDowell et~al\mbox{.}(2006)]%
        {McDowell2006-xz}
\bibfield{author}{\bibinfo{person}{Charlie McDowell}, \bibinfo{person}{Linda
  Werner}, \bibinfo{person}{Heather~E Bullock}, {and} \bibinfo{person}{Julian
  Fernald}.} \bibinfo{year}{2006}\natexlab{}.
\newblock \showarticletitle{Pair programming improves student retention,
  confidence, and program quality}.
\newblock \bibinfo{journal}{\emph{Commun. ACM}} \bibinfo{volume}{49},
  \bibinfo{number}{8} (\bibinfo{date}{Aug.} \bibinfo{year}{2006}),
  \bibinfo{pages}{90--95}.
\newblock
\showISSN{0001-0782, 1557-7317}
\urldef\tempurl%
\url{https://doi.org/10.1145/1145287.1145293}
\showDOI{\tempurl}


\bibitem[Mozannar et~al\mbox{.}(2022)]%
        {Mozannar2022-ng}
\bibfield{author}{\bibinfo{person}{Hussein Mozannar}, \bibinfo{person}{Gagan
  Bansal}, \bibinfo{person}{Adam Fourney}, {and} \bibinfo{person}{Eric
  Horvitz}.} \bibinfo{year}{2022}\natexlab{}.
\newblock \showarticletitle{Reading between the lines: Modeling user behavior
  and costs in {AI-assisted} programming}.
\newblock \bibinfo{journal}{\emph{ArXiv}} (\bibinfo{year}{2022}).
\newblock
\urldef\tempurl%
\url{https://doi.org/10.48550/ARXIV.2210.14306}
\showDOI{\tempurl}


\bibitem[Murillo and D'Angelo(2023)]%
        {Murillo2023-ld}
\bibfield{author}{\bibinfo{person}{Ambar Murillo} {and} \bibinfo{person}{Sarah
  D'Angelo}.} \bibinfo{year}{2023}\natexlab{}.
\newblock \showarticletitle{An Engineering Perspective on Writing Assistants
  for Productivity and Creative Code}.
\newblock \bibinfo{journal}{\emph{The Second Workshop on Intelligent and
  Interactive Writing Assistants}} (\bibinfo{year}{2023}).
\newblock
\urldef\tempurl%
\url{https://cdn.glitch.global/d058c114-3406-43be-8a3c-d3afff35eda2/paper1_2023.pdf}
\showURL{%
\tempurl}


\bibitem[Murphy et~al\mbox{.}(2010)]%
        {Murphy2010-mj}
\bibfield{author}{\bibinfo{person}{Laurie Murphy}, \bibinfo{person}{Sue
  Fitzgerald}, \bibinfo{person}{Brian Hanks}, {and} \bibinfo{person}{Ren{\'e}e
  McCauley}.} \bibinfo{year}{2010}\natexlab{}.
\newblock \showarticletitle{Pair debugging: a transactive discourse analysis}.
  In \bibinfo{booktitle}{\emph{Proceedings of the Sixth international workshop
  on Computing education research}} (Aarhus, Denmark)
  \emph{(\bibinfo{series}{ICER '10})}. \bibinfo{publisher}{Association for
  Computing Machinery}, \bibinfo{address}{New York, NY, USA},
  \bibinfo{pages}{51--58}.
\newblock
\showISBNx{9781450302579}
\urldef\tempurl%
\url{https://doi.org/10.1145/1839594.1839604}
\showDOI{\tempurl}


\bibitem[Myers et~al\mbox{.}(2016)]%
        {Myers2016-jk}
\bibfield{author}{\bibinfo{person}{Brad~A Myers}, \bibinfo{person}{Amy~J Ko},
  \bibinfo{person}{Thomas~D LaToza}, {and} \bibinfo{person}{Youngseok Yoon}.}
  \bibinfo{year}{2016}\natexlab{}.
\newblock \showarticletitle{Programmers are users too: Human-centered methods
  for improving programming tools}.
\newblock \bibinfo{journal}{\emph{Computer}} \bibinfo{volume}{49},
  \bibinfo{number}{7} (\bibinfo{date}{July} \bibinfo{year}{2016}),
  \bibinfo{pages}{44--52}.
\newblock
\showISSN{0018-9162, 1558-0814}
\urldef\tempurl%
\url{https://doi.org/10.1109/MC.2016.200}
\showDOI{\tempurl}


\bibitem[Nagappan et~al\mbox{.}(2003)]%
        {Nagappan2003-hb}
\bibfield{author}{\bibinfo{person}{Nachiappan Nagappan},
  \bibinfo{person}{Laurie Williams}, \bibinfo{person}{Miriam Ferzli},
  \bibinfo{person}{Eric Wiebe}, \bibinfo{person}{Kai Yang},
  \bibinfo{person}{Carol Miller}, {and} \bibinfo{person}{Suzanne Balik}.}
  \bibinfo{year}{2003}\natexlab{}.
\newblock \showarticletitle{Improving the {CS1} experience with pair
  programming}. In \bibinfo{booktitle}{\emph{Proceedings of the 34th {SIGCSE}
  technical symposium on Computer science education}} (Reno Navada USA).
  \bibinfo{publisher}{ACM}, \bibinfo{address}{New York, NY, USA}.
\newblock
\showISBNx{9781581136487}
\urldef\tempurl%
\url{https://doi.org/10.1145/611892.612006}
\showDOI{\tempurl}


\bibitem[Nguyen and Nadi(2022)]%
        {Nguyen2022-hk}
\bibfield{author}{\bibinfo{person}{N Nguyen} {and} \bibinfo{person}{Sarah
  Nadi}.} \bibinfo{year}{2022}\natexlab{}.
\newblock \showarticletitle{An Empirical Evaluation of {GitHub} Copilot's Code
  Suggestions}.
\newblock \bibinfo{journal}{\emph{2022 IEEE/ACM 19th International Conference
  on Mining Software Repositories (MSR)}} (\bibinfo{year}{2022}).
\newblock
\urldef\tempurl%
\url{https://doi.org/10.1145/3524842.3528470}
\showDOI{\tempurl}


\bibitem[Ogan et~al\mbox{.}(2012)]%
        {Ogan2012-ip}
\bibfield{author}{\bibinfo{person}{Amy Ogan}, \bibinfo{person}{Samantha
  Finkelstein}, \bibinfo{person}{Elijah Mayfield}, \bibinfo{person}{Claudia
  D'Adamo}, \bibinfo{person}{Noboru Matsuda}, {and} \bibinfo{person}{Justine
  Cassell}.} \bibinfo{year}{2012}\natexlab{}.
\newblock \showarticletitle{``Oh dear stacy!'': social interaction,
  elaboration, and learning with teachable agents}. In
  \bibinfo{booktitle}{\emph{Proceedings of the {SIGCHI} Conference on Human
  Factors in Computing Systems}} (Austin, Texas, USA)
  \emph{(\bibinfo{series}{CHI '12})}. \bibinfo{publisher}{Association for
  Computing Machinery}, \bibinfo{address}{New York, NY, USA},
  \bibinfo{pages}{39--48}.
\newblock
\showISBNx{9781450310154}
\urldef\tempurl%
\url{https://doi.org/10.1145/2207676.2207684}
\showDOI{\tempurl}


\bibitem[Padmanabhuni et~al\mbox{.}(2012)]%
        {Padmanabhuni2012-pp}
\bibfield{author}{\bibinfo{person}{Venkata Vinod~Kumar Padmanabhuni},
  \bibinfo{person}{Hari~Praveen Tadiparthi}, {and}
  \bibinfo{person}{Sagar~Madina Muralidhar~Yanamadala}.}
  \bibinfo{year}{2012}\natexlab{}.
\newblock \showarticletitle{Effective pair programming practice-an experimental
  study}.
\newblock \bibinfo{journal}{\emph{Journal of Emerging Trends in Computing and
  Information Sciences}} \bibinfo{volume}{3}, \bibinfo{number}{4}
  (\bibinfo{year}{2012}), \bibinfo{pages}{471--479}.
\newblock
\urldef\tempurl%
\url{http://www.agilemethod.csie.ncu.edu.tw/agileMethod/download/2012papers/2012%20Effective%20Pair%20Programming%20Practice-%20An%20Experimental%20Study/Effective%20Pair%20Programming%20Practice-%20An%20Experimental%20Study.pdf}
\showURL{%
\tempurl}


\bibitem[Pardos and Bhandari(2023)]%
        {Pardos2023-sb}
\bibfield{author}{\bibinfo{person}{Zachary~A Pardos} {and}
  \bibinfo{person}{Shreya Bhandari}.} \bibinfo{year}{2023}\natexlab{}.
\newblock \showarticletitle{Learning gain differences between {ChatGPT} and
  human tutor generated algebra hints}.
\newblock  (\bibinfo{date}{Feb.} \bibinfo{year}{2023}).
\newblock
\showeprint[arxiv]{2302.06871}~[cs.CY]
\urldef\tempurl%
\url{http://arxiv.org/abs/2302.06871}
\showURL{%
\tempurl}


\bibitem[Pearce et~al\mbox{.}(2021)]%
        {Pearce2021-et}
\bibfield{author}{\bibinfo{person}{Hammond Pearce}, \bibinfo{person}{Baleegh
  Ahmad}, \bibinfo{person}{Benjamin Tan}, \bibinfo{person}{Brendan
  Dolan-Gavitt}, {and} \bibinfo{person}{Ramesh Karri}.}
  \bibinfo{year}{2021}\natexlab{}.
\newblock \showarticletitle{Asleep at the Keyboard? Assessing the Security of
  {GitHub} Copilot's Code Contributions}.
\newblock  (\bibinfo{date}{Aug.} \bibinfo{year}{2021}).
\newblock
\showeprint[arxiv]{2108.09293}~[cs.CR]
\urldef\tempurl%
\url{http://arxiv.org/abs/2108.09293}
\showURL{%
\tempurl}


\bibitem[Peng et~al\mbox{.}(2023)]%
        {Peng2023-uf}
\bibfield{author}{\bibinfo{person}{Sida Peng}, \bibinfo{person}{Eirini
  Kalliamvakou}, \bibinfo{person}{Peter Cihon}, {and} \bibinfo{person}{Mert
  Demirer}.} \bibinfo{year}{2023}\natexlab{}.
\newblock \showarticletitle{The Impact of {AI} on Developer Productivity:
  Evidence from {GitHub} Copilot}.
\newblock  (\bibinfo{date}{Feb.} \bibinfo{year}{2023}).
\newblock
\showeprint[arxiv]{2302.06590}~[cs.SE]
\urldef\tempurl%
\url{http://arxiv.org/abs/2302.06590}
\showURL{%
\tempurl}


\bibitem[Perscheid et~al\mbox{.}(2017)]%
        {Perscheid2017-sw}
\bibfield{author}{\bibinfo{person}{Michael Perscheid},
  \bibinfo{person}{Benjamin Siegmund}, \bibinfo{person}{Marcel Taeumel}, {and}
  \bibinfo{person}{Robert Hirschfeld}.} \bibinfo{year}{2017}\natexlab{}.
\newblock \showarticletitle{Studying the advancement in debugging practice of
  professional software developers}.
\newblock \bibinfo{journal}{\emph{Software Quality Journal}}
  \bibinfo{volume}{25}, \bibinfo{number}{1} (\bibinfo{date}{March}
  \bibinfo{year}{2017}), \bibinfo{pages}{83--110}.
\newblock
\showISSN{1573-1367}
\urldef\tempurl%
\url{https://doi.org/10.1007/s11219-015-9294-2}
\showDOI{\tempurl}


\bibitem[Plonka et~al\mbox{.}(2011)]%
        {Plonka2011-vd}
\bibfield{author}{\bibinfo{person}{Laura Plonka}, \bibinfo{person}{Judith
  Segal}, \bibinfo{person}{Helen Sharp}, {and} \bibinfo{person}{Janet van~der
  Linden}.} \bibinfo{year}{2011}\natexlab{}.
\newblock \showarticletitle{Collaboration in Pair Programming: Driving and
  Switching}. In \bibinfo{booktitle}{\emph{Agile Processes in Software
  Engineering and Extreme Programming - 12th International Conference, {XP}
  2011, Madrid, Spain, May 10-13, 2011. Proceedings}},
  Vol.~\bibinfo{volume}{77}. \bibinfo{publisher}{unknown},
  \bibinfo{pages}{43--59}.
\newblock
\showISSN{1865-1348}
\urldef\tempurl%
\url{https://doi.org/10.1007/978-3-642-20677-1\_4}
\showDOI{\tempurl}


\bibitem[Prather et~al\mbox{.}(2023)]%
        {Prather2023-kj}
\bibfield{author}{\bibinfo{person}{James Prather}, \bibinfo{person}{Brent~N
  Reeves}, \bibinfo{person}{Paul Denny}, \bibinfo{person}{Brett~A Becker},
  \bibinfo{person}{Juho Leinonen}, \bibinfo{person}{Andrew Luxton-Reilly},
  \bibinfo{person}{Garrett Powell}, \bibinfo{person}{James Finnie-Ansley},
  {and} \bibinfo{person}{Eddie~Antonio Santos}.}
  \bibinfo{year}{2023}\natexlab{}.
\newblock \showarticletitle{``It's Weird That it Knows What {I} Want'':
  Usability and Interactions with Copilot for Novice Programmers}.
\newblock  (\bibinfo{date}{April} \bibinfo{year}{2023}).
\newblock
\showeprint[arxiv]{2304.02491}~[cs.HC]
\urldef\tempurl%
\url{http://arxiv.org/abs/2304.02491}
\showURL{%
\tempurl}


\bibitem[Preston(2006)]%
        {Preston2006-wi}
\bibfield{author}{\bibinfo{person}{David Preston}.}
  \bibinfo{year}{2006}\natexlab{}.
\newblock \showarticletitle{Using collaborative learning research to enhance
  pair programming pedagogy}.
\newblock \bibinfo{journal}{\emph{SIGITE Newsl.}} \bibinfo{volume}{3},
  \bibinfo{number}{1} (\bibinfo{date}{Jan.} \bibinfo{year}{2006}),
  \bibinfo{pages}{16--21}.
\newblock
\showISSN{2166-1685}
\urldef\tempurl%
\url{https://doi.org/10.1145/1113378.1113381}
\showDOI{\tempurl}


\bibitem[Puryear and Sprint(2022)]%
        {Puryear2022-db}
\bibfield{author}{\bibinfo{person}{Ben Puryear} {and} \bibinfo{person}{Gina
  Sprint}.} \bibinfo{year}{2022}\natexlab{}.
\newblock \showarticletitle{Github copilot in the classroom: learning to code
  with {AI} assistance}.
\newblock \bibinfo{journal}{\emph{J. Comput. Sci. Coll.}} \bibinfo{volume}{38},
  \bibinfo{number}{1} (\bibinfo{date}{Dec.} \bibinfo{year}{2022}),
  \bibinfo{pages}{37--47}.
\newblock
\showISSN{1937-4771}
\urldef\tempurl%
\url{https://dl.acm.org/doi/pdf/10.5555/3575618.3575622}
\showURL{%
\tempurl}


\bibitem[Robe and Kuttal(2022)]%
        {Robe2022-rz}
\bibfield{author}{\bibinfo{person}{Peter Robe} {and}
  \bibinfo{person}{Sandeep~Kaur Kuttal}.} \bibinfo{year}{2022}\natexlab{}.
\newblock \showarticletitle{Designing {PairBuddy---A} Conversational Agent for
  Pair Programming}.
\newblock \bibinfo{journal}{\emph{ACM Trans. Comput.-Hum. Interact.}}
  \bibinfo{volume}{29}, \bibinfo{number}{4} (\bibinfo{date}{May}
  \bibinfo{year}{2022}), \bibinfo{pages}{1--44}.
\newblock
\showISSN{1073-0516}
\urldef\tempurl%
\url{https://doi.org/10.1145/3498326}
\showDOI{\tempurl}


\bibitem[Salleh et~al\mbox{.}(2011)]%
        {Salleh2011-dd}
\bibfield{author}{\bibinfo{person}{Norsaremah Salleh}, \bibinfo{person}{Emilia
  Mendes}, {and} \bibinfo{person}{John Grundy}.}
  \bibinfo{year}{2011}\natexlab{}.
\newblock \showarticletitle{Empirical Studies of Pair Programming for {CS/SE}
  Teaching in Higher Education: A Systematic Literature Review}.
\newblock \bibinfo{journal}{\emph{IEEE Trans. Software Eng.}}
  \bibinfo{volume}{37}, \bibinfo{number}{4} (\bibinfo{date}{July}
  \bibinfo{year}{2011}), \bibinfo{pages}{509--525}.
\newblock
\showISSN{1939-3520}
\urldef\tempurl%
\url{https://doi.org/10.1109/TSE.2010.59}
\showDOI{\tempurl}


\bibitem[Sankaranarayanan et~al\mbox{.}(2020)]%
        {Sankaranarayanan2020-fc}
\bibfield{author}{\bibinfo{person}{S Sankaranarayanan}, \bibinfo{person}{S~R
  Kandimalla}, \bibinfo{person}{S Hasan}, {and} \bibinfo{person}{{others}}.}
  \bibinfo{year}{2020}\natexlab{}.
\newblock \showarticletitle{Agent-in-the-loop: Conversational agent support in
  service of reflection for learning during collaborative programming}.
\newblock \bibinfo{journal}{\emph{Artif. Intell.}} (\bibinfo{year}{2020}).
\newblock
\showISSN{0004-3702}
\urldef\tempurl%
\url{https://link.springer.com/chapter/10.1007/978-3-030-52240-7_50}
\showURL{%
\tempurl}


\bibitem[Sarkar et~al\mbox{.}(2022)]%
        {Sarkar2022-cy}
\bibfield{author}{\bibinfo{person}{Advait Sarkar}, \bibinfo{person}{Andrew~D
  Gordon}, \bibinfo{person}{Carina Negreanu}, \bibinfo{person}{Christian
  Poelitz}, \bibinfo{person}{Sruti~Srinivasa Ragavan}, {and}
  \bibinfo{person}{Ben Zorn}.} \bibinfo{year}{2022}\natexlab{}.
\newblock \showarticletitle{What is it like to program with artificial
  intelligence?}
\newblock  (\bibinfo{date}{Aug.} \bibinfo{year}{2022}).
\newblock
\showeprint[arxiv]{2208.06213}~[cs.HC]
\urldef\tempurl%
\url{http://arxiv.org/abs/2208.06213}
\showURL{%
\tempurl}


\bibitem[Schroeder et~al\mbox{.}(2013)]%
        {Schroeder2013-ub}
\bibfield{author}{\bibinfo{person}{Noah~L Schroeder},
  \bibinfo{person}{Olusola~O Adesope}, {and} \bibinfo{person}{Rachel~Barouch
  Gilbert}.} \bibinfo{year}{2013}\natexlab{}.
\newblock \showarticletitle{How Effective are Pedagogical Agents for Learning?
  A {Meta-Analytic} Review}.
\newblock \bibinfo{journal}{\emph{Journal of Educational Computing Research}}
  \bibinfo{volume}{49}, \bibinfo{number}{1} (\bibinfo{date}{July}
  \bibinfo{year}{2013}), \bibinfo{pages}{1--39}.
\newblock
\showISSN{0735-6331}
\urldef\tempurl%
\url{https://doi.org/10.2190/EC.49.1.a}
\showDOI{\tempurl}


\bibitem[Shneiderman and Maes(1997)]%
        {Shneiderman1997-gh}
\bibfield{author}{\bibinfo{person}{Ben Shneiderman} {and}
  \bibinfo{person}{Pattie Maes}.} \bibinfo{year}{1997}\natexlab{}.
\newblock \showarticletitle{Direct manipulation vs. interface agents}.
\newblock \bibinfo{journal}{\emph{Interactions}} \bibinfo{volume}{4},
  \bibinfo{number}{6} (\bibinfo{date}{Nov.} \bibinfo{year}{1997}),
  \bibinfo{pages}{42--61}.
\newblock
\showISSN{1072-5520}
\urldef\tempurl%
\url{https://doi.org/10.1145/267505.267514}
\showDOI{\tempurl}


\bibitem[Sillitti et~al\mbox{.}(2012)]%
        {Sillitti2012-ma}
\bibfield{author}{\bibinfo{person}{Alberto Sillitti},
  \bibinfo{person}{Giancarlo Succi}, {and} \bibinfo{person}{Jelena Vlasenko}.}
  \bibinfo{year}{2012}\natexlab{}.
\newblock \showarticletitle{Understanding the impact of Pair Programming on
  developers attention: A case study on a large industrial experimentation}. In
  \bibinfo{booktitle}{\emph{2012 34th International Conference on Software
  Engineering ({ICSE})}} (Zurich). \bibinfo{publisher}{IEEE},
  \bibinfo{pages}{1094--1101}.
\newblock
\showISBNx{9781467310666, 9781467310673}
\showISSN{1558-1225}
\urldef\tempurl%
\url{https://doi.org/10.1109/ICSE.2012.6227110}
\showDOI{\tempurl}


\bibitem[Sison(2009)]%
        {Sison2009-mp}
\bibfield{author}{\bibinfo{person}{Raymund Sison}.}
  \bibinfo{year}{2009}\natexlab{}.
\newblock \showarticletitle{Investigating the Effect of Pair Programming and
  Software Size on Software Quality and Programmer Productivity}. In
  \bibinfo{booktitle}{\emph{2009 16th {Asia-Pacific} Software Engineering
  Conference}}. \bibinfo{pages}{187--193}.
\newblock
\showISSN{1530-1362}
\urldef\tempurl%
\url{https://doi.org/10.1109/APSEC.2009.71}
\showDOI{\tempurl}


\bibitem[Smith et~al\mbox{.}(2012)]%
        {Smith2012-ip}
\bibfield{author}{\bibinfo{person}{Joanna Smith}, \bibinfo{person}{Joe
  Tessler}, \bibinfo{person}{Elliot Kramer}, {and} \bibinfo{person}{Calvin
  Lin}.} \bibinfo{year}{2012}\natexlab{}.
\newblock \showarticletitle{Using peer review to teach software testing}. In
  \bibinfo{booktitle}{\emph{Proceedings of the ninth annual international
  conference on International computing education research}} (Auckland, New
  Zealand) \emph{(\bibinfo{series}{ICER '12})}. \bibinfo{publisher}{Association
  for Computing Machinery}, \bibinfo{address}{New York, NY, USA},
  \bibinfo{pages}{93--98}.
\newblock
\showISBNx{9781450316040}
\urldef\tempurl%
\url{https://doi.org/10.1145/2361276.2361295}
\showDOI{\tempurl}


\bibitem[Sonkar et~al\mbox{.}(2023)]%
        {Sonkar2023-mr}
\bibfield{author}{\bibinfo{person}{Shashank Sonkar}, \bibinfo{person}{Lucy
  Liu}, \bibinfo{person}{Debshila~Basu Mallick}, {and}
  \bibinfo{person}{Richard~G Baraniuk}.} \bibinfo{year}{2023}\natexlab{}.
\newblock \showarticletitle{{CLASS} Meet {SPOCK}: An Education Tutoring Chatbot
  based on Learning Science Principles}.
\newblock  (\bibinfo{date}{May} \bibinfo{year}{2023}).
\newblock
\showeprint[arxiv]{2305.13272}~[cs.CL]
\urldef\tempurl%
\url{http://arxiv.org/abs/2305.13272}
\showURL{%
\tempurl}


\bibitem[Sun and Marakas(2009)]%
        {Sun2009-ja}
\bibfield{author}{\bibinfo{person}{W Sun} {and} \bibinfo{person}{G Marakas}.}
  \bibinfo{year}{2009}\natexlab{}.
\newblock \showarticletitle{The True Cost of Pair Programming: Development of a
  Comprehensive Model and Test}.
\newblock \bibinfo{journal}{\emph{Americas Conference on Information Systems}}
  (\bibinfo{year}{2009}).
\newblock
\urldef\tempurl%
\url{https://www.semanticscholar.org/paper/647fc48650e4f19962c8a6feb87f3bdedde9dd04}
\showURL{%
\tempurl}


\bibitem[Tan(2023)]%
        {Tan2023-wv}
\bibfield{author}{\bibinfo{person}{Chenhao Tan}.}
  \bibinfo{year}{2023}\natexlab{}.
\newblock \bibinfo{title}{On {AI} Anthropomorphism - {Human-Centered} {AI} -
  Medium}.
\newblock
  \bibinfo{howpublished}{\url{https://medium.com/human-centered-ai/on-ai-anthropomorphism-abff4cecc5ae}}.
\newblock
\urldef\tempurl%
\url{https://medium.com/human-centered-ai/on-ai-anthropomorphism-abff4cecc5ae}
\showURL{%
\tempurl}
\newblock
\shownote{Accessed: 2023-4-23}.


\bibitem[Thomas et~al\mbox{.}(2003)]%
        {Thomas2003-ud}
\bibfield{author}{\bibinfo{person}{Lynda Thomas}, \bibinfo{person}{Mark
  Ratcliffe}, {and} \bibinfo{person}{Ann Robertson}.}
  \bibinfo{year}{2003}\natexlab{}.
\newblock \showarticletitle{Code warriors and code-a-phobes: a study in
  attitude and pair programming}.
\newblock \bibinfo{journal}{\emph{SIGCSE Bull.}} \bibinfo{volume}{35},
  \bibinfo{number}{1} (\bibinfo{date}{Jan.} \bibinfo{year}{2003}),
  \bibinfo{pages}{363--367}.
\newblock
\showISSN{0097-8418}
\urldef\tempurl%
\url{https://doi.org/10.1145/792548.612007}
\showDOI{\tempurl}


\bibitem[Thorp(2023)]%
        {Thorp2023-ah}
\bibfield{author}{\bibinfo{person}{H~Holden Thorp}.}
  \bibinfo{year}{2023}\natexlab{}.
\newblock \showarticletitle{{ChatGPT} is fun, but not an author}.
\newblock \bibinfo{journal}{\emph{Science}} \bibinfo{volume}{379},
  \bibinfo{number}{6630} (\bibinfo{date}{Jan.} \bibinfo{year}{2023}),
  \bibinfo{pages}{313}.
\newblock
\showISSN{0036-8075, 1095-9203}
\urldef\tempurl%
\url{https://doi.org/10.1126/science.adg7879}
\showDOI{\tempurl}


\bibitem[Umapathy and Ritzhaupt(2017)]%
        {Umapathy2017-cd}
\bibfield{author}{\bibinfo{person}{Karthikeyan Umapathy} {and}
  \bibinfo{person}{Albert~D Ritzhaupt}.} \bibinfo{year}{2017}\natexlab{}.
\newblock \showarticletitle{A {Meta-Analysis} of {Pair-Programming} in Computer
  Programming Courses: Implications for Educational Practice}.
\newblock \bibinfo{journal}{\emph{ACM Trans. Comput. Educ.}}
  \bibinfo{volume}{17}, \bibinfo{number}{4} (\bibinfo{date}{Aug.}
  \bibinfo{year}{2017}), \bibinfo{pages}{1--13}.
\newblock
\urldef\tempurl%
\url{https://doi.org/10.1145/2996201}
\showDOI{\tempurl}


\bibitem[Vaithilingam et~al\mbox{.}(2022)]%
        {Vaithilingam2022-jh}
\bibfield{author}{\bibinfo{person}{Priyan Vaithilingam},
  \bibinfo{person}{Tianyi Zhang}, {and} \bibinfo{person}{Elena~L Glassman}.}
  \bibinfo{year}{2022}\natexlab{}.
\newblock \showarticletitle{Expectation vs. Experience: Evaluating the
  Usability of Code Generation Tools Powered by Large Language Models}. In
  \bibinfo{booktitle}{\emph{Extended Abstracts of the 2022 {CHI} Conference on
  Human Factors in Computing Systems}} (New Orleans, LA, USA)
  \emph{(\bibinfo{series}{CHI EA '22}, \bibinfo{number}{Article 332})}.
  \bibinfo{publisher}{Association for Computing Machinery},
  \bibinfo{address}{New York, NY, USA}, \bibinfo{pages}{1--7}.
\newblock
\showISBNx{9781450391566}
\urldef\tempurl%
\url{https://doi.org/10.1145/3491101.3519665}
\showDOI{\tempurl}


\bibitem[Wang et~al\mbox{.}(2022)]%
        {Wang2022-sr}
\bibfield{author}{\bibinfo{person}{Zichao Wang}, \bibinfo{person}{Jakob
  Valdez}, \bibinfo{person}{Debshila Basu~Mallick}, {and}
  \bibinfo{person}{Richard~G Baraniuk}.} \bibinfo{year}{2022}\natexlab{}.
\newblock \showarticletitle{Towards {Human-Like} Educational Question
  Generation with Large Language Models}. In
  \bibinfo{booktitle}{\emph{Artificial Intelligence in Education}}.
  \bibinfo{publisher}{Springer International Publishing},
  \bibinfo{pages}{153--166}.
\newblock
\urldef\tempurl%
\url{https://doi.org/10.1007/978-3-031-11644-5\_13}
\showDOI{\tempurl}


\bibitem[Williams and Upchurch(2001)]%
        {Williams2001-mk}
\bibfield{author}{\bibinfo{person}{Laurie Williams} {and}
  \bibinfo{person}{Richard~L Upchurch}.} \bibinfo{year}{2001}\natexlab{}.
\newblock \showarticletitle{In support of student pair-programming}. In
  \bibinfo{booktitle}{\emph{Proceedings of the thirty-second {SIGCSE} technical
  symposium on Computer Science Education}} (Charlotte North Carolina USA).
  \bibinfo{publisher}{ACM}, \bibinfo{address}{New York, NY, USA}.
\newblock
\showISBNx{9781581133295}
\urldef\tempurl%
\url{https://doi.org/10.1145/364447.364614}
\showDOI{\tempurl}


\bibitem[Williams et~al\mbox{.}(2002)]%
        {Williams2002-de}
\bibfield{author}{\bibinfo{person}{Laurie Williams}, \bibinfo{person}{Eric
  Wiebe}, \bibinfo{person}{Kai Yang}, \bibinfo{person}{Miriam Ferzli}, {and}
  \bibinfo{person}{Carol Miller}.} \bibinfo{year}{2002}\natexlab{}.
\newblock \showarticletitle{In support of pair programming in the introductory
  computer science course}.
\newblock \bibinfo{journal}{\emph{Comput. Sci. Educ.}} \bibinfo{volume}{12},
  \bibinfo{number}{3} (\bibinfo{date}{Sept.} \bibinfo{year}{2002}),
  \bibinfo{pages}{197--212}.
\newblock
\showISSN{0899-3408, 1744-5175}
\urldef\tempurl%
\url{https://doi.org/10.1076/csed.12.3.197.8618}
\showDOI{\tempurl}


\bibitem[Wong et~al\mbox{.}(2022)]%
        {Wong2022-dt}
\bibfield{author}{\bibinfo{person}{Dakota Wong}, \bibinfo{person}{Austin
  Kothig}, {and} \bibinfo{person}{Patrick Lam}.}
  \bibinfo{year}{2022}\natexlab{}.
\newblock \showarticletitle{Exploring the Verifiability of Code Generated by
  {GitHub} Copilot}.
\newblock \bibinfo{journal}{\emph{ACM on Programming Languages}}
  (\bibinfo{year}{2022}).
\newblock
\urldef\tempurl%
\url{https://www.semanticscholar.org/paper/b5051fedaf17836f6b2a042cc4af4155159795c5}
\showURL{%
\tempurl}


\bibitem[Yeti{\c s}tiren et~al\mbox{.}(2022)]%
        {Yetistiren2022-so}
\bibfield{author}{\bibinfo{person}{Burak Yeti{\c s}tiren},
  \bibinfo{person}{I{\c s}ik {\"O}zsoy}, {and} \bibinfo{person}{Eray
  T{\"u}z{\"u}n}.} \bibinfo{year}{2022}\natexlab{}.
\newblock \showarticletitle{Assessing the Quality of {GitHub} Copilot's Code
  Generation}. In \bibinfo{booktitle}{\emph{18th International Conference on
  Predictive Models and Data Analytics in Software Engineering ({PROMISE}
  '22)}}.
\newblock
\urldef\tempurl%
\url{https://doi.org/10.1145/3558489.3559072}
\showDOI{\tempurl}


\bibitem[Ziegler et~al\mbox{.}(2022)]%
        {Ziegler2022-fz}
\bibfield{author}{\bibinfo{person}{Albert Ziegler}, \bibinfo{person}{Eirini
  Kalliamvakou}, \bibinfo{person}{X~Alice Li}, \bibinfo{person}{Andrew Rice},
  \bibinfo{person}{Devon Rifkin}, \bibinfo{person}{Shawn Simister},
  \bibinfo{person}{Ganesh Sittampalam}, {and} \bibinfo{person}{Edward
  Aftandilian}.} \bibinfo{year}{2022}\natexlab{}.
\newblock \showarticletitle{Productivity assessment of neural code completion}.
  In \bibinfo{booktitle}{\emph{Proceedings of the 6th {ACM} {SIGPLAN}
  International Symposium on Machine Programming}} (San Diego CA USA).
  \bibinfo{publisher}{ACM}, \bibinfo{address}{New York, NY, USA}.
\newblock
\urldef\tempurl%
\url{https://doi.org/10.1145/3520312.3534864}
\showDOI{\tempurl}


\end{thebibliography}
\bibliographystyle{ACM-Reference-Format}

\end{document}